\documentclass[12pt,preprint]{aastex}
\newcommand{\kms}{km~s$^{-1}$}

\shortauthors{J{\o}rgensen et al.}

\begin{document}

\title{PROSAC: A Submillimeter Array Survey of Low-Mass Protostars\\
  I. Overview of Program: Envelopes, Disks, Outflows and Hot Cores}

\author{Jes K. J{\o}rgensen\altaffilmark{1}, Tyler L. Bourke\altaffilmark{1}, Philip C. Myers\altaffilmark{1}, James Di Francesco\altaffilmark{2}, Ewine F. van Dishoeck\altaffilmark{3}, Chin-Fei Lee\altaffilmark{4}, Nagayoshi Ohashi\altaffilmark{5}, Fredrik L. Sch\"{o}ier\altaffilmark{6}, Shigehisa Takakuwa\altaffilmark{7}, David J. Wilner\altaffilmark{1}, \& Qizhou Zhang\altaffilmark{1}}
\altaffiltext{1}{Harvard-Smithsonian Center for Astrophysics, 60 Garden Street MS42, Cambridge, MA 02138, USA {\tt jjorgensen@cfa.harvard.edu}}
\altaffiltext{2}{Herzberg Institute of Astrophysics, National Research Council of Canada, 5071 West Saanich Road, Victoria, BC V9E 2E7, Canada}
\altaffiltext{3}{Leiden Observatory, P.O. Box 9513, NL-2300 RA Leiden, The Netherlands}
\altaffiltext{4}{Harvard-Smithsonian Center for Astrophysics, Submillimeter Array Project, 645 North A'ohoku Place, Hilo, HI 96720, USA}
\altaffiltext{5}{Academia Sinica Institute of Astronomy and Astrophysics, P.O. Box 23-141, Taipei 106, Taiwan}
\altaffiltext{6}{Stockholm Observatory, AlbaNova, SE-106 91, Stockholm, Sweden}
\altaffiltext{7}{National Astronomical Observatory of Japan, ALMA Project Office, Osawa 2-21-1, Mitaka, Tokyo, 181-8588, Japan}

\begin{abstract}
  This paper presents a large spectral line and continuum survey of 8
  deeply embedded, low-mass protostellar cores using the Submillimeter
  Array. Each source was observed with three different spectral
  settings, to include high excitation lines of some of the most
  common molecular species, CO, HCO$^+$, CS, SO, H$_2$CO, CH$_3$OH and
  SiO. Line emission from 11 molecular species (including
  isotopologues) originating from warm and dense gas have been imaged
  at high angular resolution (1--3$''$; typically corresponding to
  200--600~AU scales) together with continuum emission at 230~GHz
  (1.3~mm) and 345~GHz (0.8~mm). Compact continuum emission is
  observed for all sources which likely originates in marginally
  optically thick circumstellar disks, with typical lower limits to
  their masses of $0.1$~$M_\odot$ (1--10\% of the masses of their
  envelopes) and having a dust opacity law, $\kappa_\nu \propto
  \nu^\beta$, with $\beta \approx 1$. Prominent outflows are present
  in CO 2--1 observations in all sources, extending over most of the
  interferometer field of view. Most outflows are highly
  collimated. Significant differences are seen in their morphologies,
  however, with some showing more jet-like structure and others
  seemingly tracing material in the outflow cavity walls. The most
  diffuse outflows are found in the sources with the lowest ratios of
  disk-to-envelope mass, and it is suggested that these sources are in
  a phase where accretion of matter from the envelope has almost
  finished and the remainder of the envelope material is being
  dispersed by the outflows. Other characteristic dynamical signatures
  are found with inverse P Cygni profiles indicative of infalling
  motions seen in the $^{13}$CO 2--1 lines toward NGC~1333-IRAS4A and
  NGC~1333-IRAS4B. Outflow-induced shocks are present on all scales in
  the protostellar environments and are most clearly traced by the
  emission of CH$_3$OH in NGC~1333-IRAS4A and NGC~1333-IRAS4B. These
  observations suggest that the emission of CH$_3$OH and H$_2$CO from
  these proposed ``hot corinos'' are related to the shocks caused by
  the protostellar outflows. One source, NGC~1333-IRAS2A, stands out
  as the only one remaining with evidence for hot, compact CH$_3$OH
  emission coincident with the embedded protostar.
\end{abstract}


\section{Introduction}
Low-mass stars form from the gravitational collapse of dense molecular
cloud cores. In the earliest stages, the young protostar is deeply
embedded in a cold envelope of infalling gas \citep[the so-called
``Class 0'' stage;][]{andre93,andreppiv}. Studies of the deeply
embedded stages reveal the properties of the youngest stellar objects
during or just after formation of the star. An interesting interplay
takes place in these earliest stages between material falling toward
the central star-disk system and that being dispersed by a
protostellar outflow while simultaneously being heated by radiation
from the central star. This paper presents a large survey of 8 deeply
embedded protostars with 1--3\arcsec\ ($\sim 200-600$~AU) resolution
in a wide range of line and continuum emission at submillimeter
wavelengths using the Submillimeter Array (SMA). These observations
are ideally suited to probe the inner warm and dense material
surrounding low-mass protostars, in contrast with lower frequency data
which are more sensitive to the cold outer envelope.

Combined studies of the physical and chemical structure of low-mass
protostars are important for a complete picture of the evolution of
young stellar objects, for understanding their structure on few
hundred AU scales and for addressing a number of questions about
low-mass star formation. For example, do all deeply embedded
protostars have disks?  How massive are these disks, both compared to
their envelopes and to the disks around the more evolved T Tauri
stars? Is there evidence that their physical properties and
composition (e.g., the properties of the dust of which they are made)
are different from those of their more evolved counterparts? What are
the time scales for the dissipation of the protostellar envelopes? How
important are protostellar outflows for the dissipation, e.g.,
compared to the amount of material accreted onto the central star+disk
system? Likewise, what is the importance of shocks associated with the
outflows in regulating the chemistry in the protostellar envelopes?
Are complex organic molecules common in these envelopes and do they
originate in shocks or due to heating by the central protostar?

Observations at (sub)millimeter wavelengths can address a number of
these issues: thermal dust continuum emission can be used to constrain
the envelope density and temperature structure whereas line
observations give information about the dynamical structure (e.g., of
the infalling material), of excitation conditions and of the
chemistry.  Submillimeter dust emission and high-excitation lines
constrain the properties of the actual protostellar envelope and are
less sensitive to the surrounding cloud material than, e.g.,
observations from high-resolution 3~mm studies
\citep[e.g.,][]{gueth97,ohashi97,hogerheijde99,looney00,difrancesco01,harvey03,n1333i2art}. Previously
such submillimeter observations were only feasible with single-dish
telescopes with resolutions of a few thousand AU for nearby star
forming regions \citep[see, e.g.,][for a recent
review]{vandishoeck05}. With the Submillimeter Array \citep[][]{ho04},
however, a new window has opened up, allowing for detailed studies of
the radial variations of both the physical/dynamical and chemical
structure of protostellar cores.

The SMA is ideal for studies of these inner regions for a number of
reasons: First, previous interferometric studies based on lower
excitation lines did not probe deeply inside the envelope since these
lines are sensitive to the chemistry of the outer, cold regions and
become optically thick. In the 325--365~GHz window a wealth of
molecular transitions constrain the chemistry in the denser ($\sim
10^7-10^8$~cm$^{-3}$) and warmer ($\sim 50-100$~K) material of the
envelope. Second, since the dust continuum flux scales with frequency
as $\nu^2$ or steeper, submillimeter observations are well suited to
probe the dust in protostellar disks. At the same time, the innermost
regions of the envelopes where the temperature increases above 100~K
are heavily diluted in a single-dish beam ($<2''$ size compared to
typical single-dish beam sizes of 10-20$''$). Third, interpretation of
the line emission from these regions relies on extrapolation of the
density and temperature distribution from observations on larger
scales.  Typical SMA observations resolve the emission down to these
scales and make it possible to disentangle the emission from the
envelope and circumstellar disk.

A number of recent papers have presented studies of individual
protostars with the SMA that focused on different aspects of the
physical and chemical structure of protostellar envelopes, disks and
outflows
\citep{chandler05,iras2sma,kuan04,lee06,palau06,takakuwa04,takakuwa06}. To
make statistical comparisons and general statements about the deeply
embedded protostars as a whole, however, the systematic effort
described here is warranted. The purpose of this paper is to describe
the details of the observations and present an overview of the results
with a few pointers to important implications.  The observational
details, including the sample, observed settings, reduction and
calibration strategy, are presented and discussed in \S\ref{obs}. In
\S\ref{contresults}, the results of the continuum observations are
discussed with simple estimates of the compact emission from the
central disks, their masses and the slope of the dust opacity law. The
line observations are discussed in \S\ref{lineresults}, showing all
observed maps and addressing which molecular species are particularly
useful for probing different aspects of the protostellar cores, and
focusing on the dynamical and chemical impact of the protostellar
outflows. Finally, \S\ref{summary} concludes the paper. This paper
serves as a starting point for a series of focused papers that will
describe more specific topics in detail and will serve as an important
reference for planning future SMA (and eventual ALMA) observations of
low-mass protostars.

\section{Observations and data reduction}\label{obs}
\subsection{Sample}
The sources observed in this program were selected from a large
single-dish survey of submillimeter continuum and line emission toward
low-mass protostars \citep{jorgensen02,paperii}. That sample consisted
of nearby ($d< 400$~pc) Class~0 objects with luminosities less than
$30~L_\odot$ - all visible from Mauna Kea (i.e., predominantly
Northern sources). For this study we selected a subset of 7 objects
that, in addition to the above single-dish survey, have been studied
extensively through aperture synthesis observations at 3~millimeter
wavelengths. To this we added the Class 0 object B335 which likewise
has been studied in great detail using (sub)millimeter single-dish
telescopes \citep{huard99,shirley00,shirley02,evans05b335}, through
longer wavelength aperture synthesis observations
\citep{wilner00,harvey03}, at near-IR wavelengths \citep{harvey01},
and using detailed dust and line radiative transfer
\citep{shirley02,evans05b335}. Detailed line and continuum radiative
transfer models exist for each of the objects which can be used to
address the extended emission associated with their envelopes
\citep[e.g.,][]{n1333i2art,iras2sma,l483art,hotcorepaper} and in
addition single-dish maps have been obtained for a subset of the lines
which will be incorporated in forthcoming papers. The full sample of
objects is summarized in Table~\ref{sample}. In
Appendix~\ref{sampledescription}, we provide a brief description of
the characteristics of each object from previous observations.

\begin{table}
\caption{Sample of sources.}\label{sample}
\begin{tabular}{llllllll}\hline\hline
Source             &        & \multicolumn{2}{c}{Pointing center\tablenotemark{a}} & Association & Distance \\
(full name)  & (short name) & $\alpha(J2000)$ & $\delta(J2000)$ &           & [pc]   \\ \hline
L1448-C(N)\tablenotemark{b}    & L1448  & 03:25:38.80  & $+$30:44:05.0 & Perseus            & 220 \\
NGC~1333-IRAS2A    & IRAS2A  & 03:28:55.70  & $+$31:14:37.0 & Perseus            & 220 \\
NGC~1333-IRAS4A    & IRAS4A & 03:29:10.50  & $+$31:13:31.0 & Perseus            & 220 \\
NGC~1333-IRAS4B    & IRAS4B & 03:29:12.00  & $+$31:13:08.0 & Perseus            & 220 \\
L1527-IRS          & L1527  & 04:39:53.90  & $+$26:03:10.0 & Taurus             & 140 \\
L483-FIR           & L483   & 18:27:29.85  & $-$04:39:38.8 & Isolated           & 200 \\
B335               & B335   & 19:37:00.90  & $+$07:34:10.0 & Isolated           & 250 \\
L1157-mm           & L1157  & 20:39:06.20  & $+$68:02:15.9 & Isolated           & 325 \\ \hline
\end{tabular}

\tablenotetext{a}{Accurate positions from fits to 230~GHz and 345~GHz continuum
observations given in Table~\ref{continuum_gauss}.} 
\tablenotetext{b}{Refers to the
known source from high angular resolution millimeter observations, often
referred to as L1448-mm or L1448-C. Recent Spitzer observations
\citep{perspitz} find a second source (in the same SCUBA core) about 8\arcsec\ south of this position.}
\end{table}

\subsection{Spectral setups}
We selected three spectral setups per source covering a wide range of
lines at 0.8~mm and 1.3~mm (345~GHz and 230~GHz) while at the same
time providing continuum observations of each core at both
wavelengths. The lines observed include a full suite of CO isotopic
lines (probing different aspects of the physical structure of the
envelopes and the associated outflows together with molecular
freeze-out), CS, C$^{34}$S, and H$^{13}$CO$^+$ (typically probing
dense gas in the envelopes), SiO and SO (which are likely shock
tracers) and CH$_3$OH and H$_2$CO (which either trace hot material in
the inner envelopes or shocked gas in the
outflows). Tables~\ref{spectral_first}--\ref{spectral_last} summarize
the observed spectral setups.

The SMA correlator covers 2~GHz bandwidth in each of the two sidebands
separated by 10~GHz. Each band is divided into 24 ``chunks'' of
104~MHz width which can be covered by varying spectral resolution. For
each spectral setting, the correlator was configured to maximize the
spectral resolution on the key species and explore the dynamical
structure of the protostars. With 128--512 channels per chunk, the
spectral resolution ranges from 0.2~\kms\ to 1~km~s$^{-1}$ with most
of the dense gas (i.e., envelope) tracers having the highest spectral
resolution of 0.2--0.3~\kms\ while the outflow tracers typically have
lower resolution of 0.5--1~\kms\
(Tables~\ref{spectral_first}--\ref{spectral_last}). The remaining
chunks were covered by 16 channels each and used to measure the
continuum typically over 1.5--1.7~GHz bandwidth in each of the two
sidebands. None of the remaining chunks are expected to include bright
lines which could affect the continuum measurements.

\begin{deluxetable}{llllll}
\tablecaption{Spectral setup: 230 GHz setting (219.43~GHz--221.43~GHz \& 229.43~GHz--231.43~GHz). \label{spectral_first}} 
\tablehead{\colhead{Chunk} & \colhead{Frequency range} & \colhead{Channels} & \colhead{Resolution} & \colhead{Lines}\tablenotemark{a} & \colhead{Frequency} \\ 
\colhead{} & \colhead{[GHz]} & \colhead{} & \colhead{[\kms]} & \colhead{} & \colhead{[GHz]} } \startdata
\multicolumn{6}{c}{\emph{LSB}} \\ \tableline
s23 & 219.506--219.610 & 512 & 0.28 & C$^{18}$O 2--1  & 219.560 \\ 
s18 & 219.909--220.013 & 256 & 0.55 & SO $5_6-4_5$   & 219.949 \\ 
s14 & 220.225--220.329 & 128 & 1.1  & $\ldots$      & $\ldots$ \\ 
s13 & 220.238--220.342 & 512 & 0.28 & $^{13}$CO 2--1 & 220.399 \\ \tableline
\multicolumn{6}{c}{\emph{USB}} \\ \tableline
s13 & 230.407--230.511 & 512 & 0.26 & $\ldots$      & $\ldots$ \\ 
s14 & 230.489--230.593 & 128 & 1.1  & CO 2--1       & 230.538 \\ 
s18 & 230.817--230.921 & 256 & 0.53 & $\ldots$      & $\ldots$ \\
s23 & 231.221--231.325 & 512 & 0.26 & (CH$_3$OH $10_{2,9}-9_{3,6}$ (A))\tablenotemark{b} & (231.281) \\ 
\enddata
\tablenotetext{a}{{Lines} indicated in parenthesis indicate
  non-detections in the given window.}  \tablenotetext{b}{Tentative
  detection in shock associated with IRAS4B outflow.}
\end{deluxetable}

\begin{deluxetable}{llllll}
\tablecaption{Spectral setup: 337 GHz setting (336.83~GHz--338.83~GHz \& 346.83~GHz--348.83~GHz).}
\tablehead{\colhead{Chunk} & \colhead{Frequency range} & \colhead{Channels} & \colhead{Resolution} & \colhead{Lines\tablenotemark{a}} & \colhead{Frequency} \\
\colhead{} & \colhead{[GHz]} & \colhead{} & \colhead{[\kms]} & \colhead{} & \colhead{[GHz]} }
\startdata
\multicolumn{6}{c}{\emph{LSB}} \\ \tableline
s22 & 337.014--337.118  & 512 & 0.18 & C$^{17}$O 3--2                      & 337.061 \\
s18 & 337.342--337.446  & 512 & 0.18 & C$^{34}$S 7--6                      & 337.397 \\
s06 & 338.326--338.430  & 128 & 0.72 & CH$_3$OH $7_{-1}-6_{-1}$ (E)        & 338.345 \\
    &                   &     &      & CH$_3$OH $7_0-6_0$                  & 338.409 \\
s05 & 338.408--338.512  & 128 & 0.72 & CH3OH $7_k-6_k$                     & $\ldots$\tablenotemark{b} \\
s02 & 338.654--338.758  & 512 & 0.18 & CH$_3$OH $7_{\pm 2}-6_{\pm 2}$ (E)  & 338.722 \\ \tableline
\multicolumn{6}{c}{\emph{USB}} \\ \tableline
s02 & 346.937--347.041  & 512 & 0.18 & H$^{13}$CO$^+$ 4--3                 & 346.998       \\
s05 & 347.183--347.287  & 128 & 0.70 & $\ldots$                            & $\ldots$      \\
s06 & 347.266--347.369  & 128 & 0.70 & SiO 8--7                            & 347.331       \\
s18 & 348.249--348.353  & 512 & 0.18 & (HN$^{13}$C 4-3)                    & (348.340)     \\
s22 & 348.577--348.681  & 512 & 0.18 & SO$_2$ $5_{3,3}-6_{0,6}$              & (348.633)     \\
\enddata
\tablenotetext{a}{{Lines} indicated in parenthesis indicate
  non-detections in the given window.}  \tablenotetext{b}{Multiple
  transitions covered (see Figure~\ref{ch3oh_spectra}).}
\end{deluxetable}

\begin{deluxetable}{llllll}
\tablecaption{Spectral setup: 342 GHz setting (340.93~GHz--342.93~GHz \& 350.93~GHz--352.93~GHz).\label{spectral_last}}
\tablehead{\colhead{Chunk} & \colhead{Frequency range} & \colhead{Channels} & \colhead{Resolution} & \colhead{Lines\tablenotemark{a}} & \colhead{Frequency} \\ 
\colhead{} & \colhead{[GHz]} & \colhead{} & \colhead{[\kms]} & \colhead{} & \colhead{[GHz]} }
\startdata
\multicolumn{6}{c}{\emph{LSB}} \\ \tableline
s10 & 342.090--342.194  & 512 & 0.18 & $\ldots$                        & $\ldots$     \\
s04 & 342.588--342.692  & 256 & 0.35 & (CH$_3$CHO $12_{3,10}-22_{1,11}$) & (342.641)    \\
s01 & 342.828--342.932  & 512 & 0.18 & CS 7--6                         & 342.883      \\ \tableline
\multicolumn{6}{c}{\emph{USB}} \\ \tableline
s01 & 350.947--351.051  & 512 & 0.17 & (NO 4--3)                       & (351.043)    \\
s04 & 351.187--351.291  & 256 & 0.35 & (SO$_2$ $5_{3,3}-4_{2,2}$)         & (351.257)    \\
s10 & 351.685--351.789  & 512 & 0.17 & H$_2$CO $5_{1,5}-4_{1,4}$         & 351.769      \\
\enddata
\tablenotetext{a}{{Lines} indicated in parenthesis indicate
  non-detections in the given window.}
\end{deluxetable}

\subsection{Observational details}
The sources were observed from November 2004 through August 2005 with
one track repeated in December 2005. For each observation, at least 6
of the 8 antennas were available and showed fringes. The November 2004
and December 2004 observations were performed with the array in the
Compact-North configuration - a configuration optimized for equatorial
sources. The remaining observations were done with the array in the
regular Compact configuration. The Compact-North configuration
includes somewhat longer baselines and provides a higher resolution of
1--1.5$''$ at 345~GHz than the Compact configuration (2--2.5$''$
resolution at 345~GHz). On the other hand the shortest baselines of
the Compact configuration are $\approx 8$~k$\lambda$ at 345~GHz
compared to 10--12~k$\lambda$ for the Compact-North configuration and
thus provide better sensitivity to extended emission.

For the sources in Perseus, two objects were observed per track. These
sources are all so close to each other that the same gain calibrators
can be used for any pair. Furthermore, they are expected to be the
brightest of our sample, so the effect of the $\sqrt{2}$ loss in
sensitivity due to less integration time is acceptable. For the
remainder of the sources, full tracks were done. For a few tracks,
only part of the data (more than 2/3 in all cases) were usable due to
weather and/or instrument related problems.

\begin{deluxetable}{lllll}
\tablecaption{Observations Log}
\tablehead{\colhead{Date} & \colhead{Source(s)} & \colhead{Setting} & \colhead{Baselines / antennas\tablenotemark{a}} & \colhead{$\tau_{\rm 225 GHz}$\tablenotemark{b}} \\ 
\colhead{} & \colhead{} & \colhead{[GHz]} & \colhead{[k$\lambda$]} & \colhead{}}
\startdata
\multicolumn{5}{c}{2004 ; Compact-North Configuration} \\ \tableline
2004-11-06\tablenotemark{c} & IRAS4A, IRAS4B & 230               & 10--107 (7)      & $0.2-0.35$           \\
2004-11-07 & L1448, IRAS2A   & 230               & 10--107 (7)      & $0.12-0.2$    \\
2004-11-08 & L1527          & 230               & 11--108 (7)      & $0.13-0.18$   \\
2004-11-11 & IRAS2A, IRAS4A  & 342               & 14--163 (7)      & $0.08-0.09$   \\
2004-11-20 & IRAS4A, IRAS4B & 337               & 15--161 (6)      & $0.055-0.075$ \\
2004-11-21 & L1448, IRAS2A   & 337               & 14--161 (6)      & $0.08-0.11$   \\
2004-11-22 & IRAS4A, IRAS4B & 230               & 10--107 (6)      & $0.1-0.19$    \\
2004-12-14\tablenotemark{d} & L1448          & 342               & 10--81\phantom{1}  (4)  & $0.06-0.07$   \\
2004-12-17 & L1527          & 337               & 13--162 (6)    & $0.03$        \\
2004-12-18 & L1527          & 342               & 15--164 (6)    & $0.04-0.07$   \\ \tableline

\multicolumn{5}{c}{2005 ; Compact Configuration} \\ \tableline
2005-06-08 & B335           & 337               & 12--81\phantom{1}  (6)           & $0.04-0.06$   \\
2005-06-14 & B335           & 342               & 12--82\phantom{1}  (7)             & $0.065-0.08$  \\
2005-06-18 & L483           & 337               & \phantom{1}7--80\phantom{1}  (7)             & $0.04-0.09$   \\
2005-06-24 & B335           & 230               & \phantom{1}5--54\phantom{1}  (7)             & $0.14-0.20$   \\
2005-07-03 & L1157          & 337               & \phantom{1}9--79\phantom{1}  (7)             & $0.08-0.1$    \\
2005-07-06 & L1157          & 230               & \phantom{1}5--53\phantom{1}  (7)             & $0.07-0.1$    \\
2005-07-10 & L483           & 342               & \phantom{1}8--82\phantom{1}  (7)             & $0.05-0.09$   \\
2005-07-24 & L483           & 230               & \phantom{1}8--54\phantom{1}  (7)             & $0.08$        \\
2005-08-22 & L1157          & 342               & \phantom{1}7--80\phantom{1}  (7)             & $0.065-0.09$  \\
2005-12-03 & IRAS4B, L1448  & 342               & 10--82\phantom{1}  (7)             & $0.06-0.07$   \\
\enddata

\tablenotetext{a}{Range of projected baseline lengths and number of
  antennas in the array providing usable data (in parenthesis).}
\tablenotetext{b}{As reported by observers} \tablenotetext{c}{Repeated
  on 2004-11-22 due to poor weather} \tablenotetext{d}{Repeated on
  2005-12-03 due to poor $(u,v)$ coverage}
\end{deluxetable}

\subsection{Data reduction}
The data were reduced in the standard way using the MIR package
\citep{qimir}. This package was developed for reduction of SMA data
based on the Owens Valley Radio Observatory MMA package
\citep{scoville93}. During the reduction, integrations with clearly
deviating phases and/or amplitudes were flagged. The passband
(spectral response) was calibrated through observations of available
planets and strong quasars (3c454.3, in particular) at the beginning
and end of each track. The complex gains were calibrated through
frequent observations of strong ($\gtrsim 1.5$~Jy) quasars relatively
close to each source (within 5--25\arcdeg). The quasar fluxes were
bootstrapped using observations of Uranus and, for a few tracks,
Callisto. Table~\ref{quasars} lists the observed quasars and their
fluxes for the individual tracks. Comparison of the derived quasar
fluxes to other measurements close-by in time suggests that a
conservative estimate of the absolute calibration accuracy is about
$\pm 30$\%, and is likely better for the most recent data due to
improvements in instrument performance and stability.

\begin{deluxetable}{lllll}
\tablecaption{Quasar fluxes in Jy (November--December 2004 and December 2005)\label{quasars}}
\tablehead{\colhead{Date} & \colhead{3C84} & \colhead{J0359+509} & \colhead{3C111} & \colhead{J0510+180} }
\startdata
\multicolumn{5}{c}{337 \& 342 GHz observations} \\ \tableline
2004-10-17 (354 GHz)\tablenotemark{a} & 1.8     & 1.8     & $\ldots$ & $\ldots$ \\
2004-11-11 (342 GHz)                  & 2.2     & 2.0     & $\ldots$ & $\ldots$ \\
2004-11-20 (337 GHz)                  & 2.1     & 1.7     & $\ldots$ & $\ldots$ \\
2004-11-21 (337 GHz)                  & 2.4     & 2.0     & $\ldots$ & $\ldots$ \\
2004-12-14 (342 GHz)                  & 2.3     & 1.9     & $\ldots$ & $\ldots$ \\
2004-12-17 (337 GHz)\tablenotemark{b} & $\ldots$ & 2.0     & 2.5     & 1.4     \\
2004-12-18 (342 GHz)\tablenotemark{b} & $\ldots$ & 1.4     & 1.7     & 1.0     \\
2005-12-03 (342 GHz)                  & 2.4     & $\ldots$ & 3.0     & $\ldots$ \\ \tableline
\multicolumn{5}{c}{230 GHz observations} \\ \tableline
2004-11-06                            & 2.7     & 2.9     & $\ldots$ & $\ldots$ \\
2004-11-07                            & 2.6     & 2.6     & $\ldots$ & $\ldots$ \\
2004-11-08                            & $\ldots$ & $\ldots$ & 3.3     & 1.4     \\
2004-11-22                            & 3.0     & 2.8     & $\ldots$ & $\ldots$ \\
\enddata
\tablenotetext{a}{Observations discussed in \cite{iras2sma}.}

\tablenotetext{b}{This calibration seems unreliable with 40\%
  difference between two subsequent nights. The J0359+509 data from the
  night of 2004-12-14 (not used for anything else) suggest the higher
  value. In both tracks on 2004-12-17 and 2004-12-18 J0359+509 was
  observed as a transit source. The ratios between the sources are
  consistent within the uncertainties. We have adopted the quasar
  fluxes from 2004-12-17 in the reduction of both tracks.}
\end{deluxetable}

\begin{deluxetable}{llllllll}
\tablecaption{Quasar fluxes in Jy (June--August 2005)}
\tablehead{\colhead{Date} & \colhead{J1751+096} & \colhead{J2148+069} & \colhead{J2202+422}   & \colhead{J1642+689}  & \colhead{J1743-038}   & \colhead{J1924-292} }
\startdata
337 GHz                   & 1.3               & 1.3               & 3.8                 &  1.3               & 2.3                 & 5.2        \\
342 GHz                   & 1.8               & 1.9               & 3.4                 &  $\ldots$          & 1.6                 & 3.7        \\
230 GHz                   & 2.0               & 2.6               & 4.6                 &  $\ldots$          & 1.2                 & 3.0        \\
\enddata
\end{deluxetable}

Each continuum map was deconvolved down to the theoretical
RMS noise level using the Miriad Clean routine. The restored maps had
RMS values of 2--40~mJy~beam$^{-1}$ (230~GHz) and
4--20~mJy~beam$^{-1}$ (345~GHz). This large range of RMS values does
not reflect the actual thermal noise level, but rather the dynamical
range of the images. For the continuum observations of IRAS4A and
IRAS4B and the CO line observations in general, significant side lobes
are seen even after the deconvolution of the maps. Slightly better
results are obtained for the strong CO lines and continuum maps using
uniform weighting which suppresses side lobes over the entire primary
beam field of view and thus improves the dynamic range, even though
the theoretical thermal noise level is slightly higher. The RMS noise
levels and beam sizes from uniform and natural weighting are
summarized in Table~\ref{rms}.
\begin{table}
\caption{RMS noise levels and beam sizes from continuum observations.}\label{rms}
\begin{tabular}{lllllll}\hline\hline
       & \multicolumn{3}{l}{Uniform weighting} & \multicolumn{3}{l}{Natural weighting} \\
Source\phantom{xxx} & RMS$_{\rm th}$\tablenotemark{a} & RMS$_{\rm cm}$\tablenotemark{b} & Beam\tablenotemark{c}          & RMS$_{\rm th}$\tablenotemark{a} & RMS$_{\rm cm}$\tablenotemark{b} & Beam\tablenotemark{c} \\
       & [mJy]           & [mJy]           &  & [mJy] & [mJy] &      \\ \hline
\multicolumn{7}{c}{230 GHz} \\ \hline
L1448  & 2.2             & 3.0             & 2.5$''$$\times$1.2$''$\phantom{0} ($+$77$^\circ$) & 1.5             & 2.7             & 2.7$''$$\times$1.7$''$\phantom{0} ($+$82$^\circ$)  \\ 
IRAS2A  & 2.2             & 3.3             & 2.5$''$$\times$1.2$''$\phantom{0} ($+$78$^\circ$) & 1.5             & 3.3             & 2.7$''$$\times$1.7$''$\phantom{0} ($+$81$^\circ$)  \\ 
IRAS4A & 2.6             & 14              & 2.5$''$$\times$1.1$''$\phantom{0} ($+$80$^\circ$) & 1.7             & 19              & 2.7$''$$\times$1.6$''$\phantom{0} ($+$86$^\circ$)  \\ 
IRAS4B & 2.5             & 23              & 2.5$''$$\times$1.1$''$\phantom{0} ($+$79$^\circ$) & 1.7             & 43              & 2.7$''$$\times$1.6$''$\phantom{0} ($+$85$^\circ$)  \\ 
L1527  & 1.5             & 1.9             & 2.6$''$$\times$1.1$''$\phantom{0} ($+$80$^\circ$) & 1.0             & 1.8             & 2.7$''$$\times$1.7$''$\phantom{0} ($+$84$^\circ$)  \\ 
L483   & 2.5             & 2.4             & 3.3$''$$\times$2.7$''$\phantom{0} ($-$87$^\circ$) & 1.4             & 2.1             & 3.9$''$$\times$3.3$''$\phantom{0} ($+$81$^\circ$)  \\ 
B335   & 2.4             & 4.5             & 3.2$''$$\times$2.9$''$\phantom{0} ($+$65$^\circ$) & 1.3             & 5.6             & 4.0$''$$\times$3.1$''$\phantom{0} ($+$41$^\circ$)  \\ 
L1157  & 3.1             & 5.7             & 4.0$''$$\times$2.4$''$\phantom{0} ($-$45$^\circ$) & 1.9             & 7.0             & 4.5$''$$\times$3.3$''$\phantom{0} ($-$50$^\circ$) \\ \hline
\multicolumn{7}{c}{345 GHz} \\ \hline                                                                                                                              
L1448  & 6.0             & 8.3             & 1.7$''$$\times$1.1$''$\phantom{0} ($+$79$^\circ$) & 4.1             & 7.3             & 2.1$''$$\times$1.7$''$\phantom{0} ($-$83$^\circ$) \\ 
IRAS2A  & 5.6             & 6.8             & 1.5$''$$\times$0.71$''$ ($+$78$^\circ$)           & 3.6             & 7.7             & 1.7$''$$\times$1.2$''$\phantom{0} ($-$83$^\circ$) \\ 
IRAS4A & 4.6             & 20              & 1.5$''$$\times$0.71$''$ ($+$81$^\circ$)           & 3.1             & 17              & 1.7$''$$\times$1.2$''$\phantom{0} ($-$78$^\circ$) \\ 
IRAS4B & 4.6             & 20              & 1.7$''$$\times$1.0$''$\phantom{0} ($+$81$^\circ$) & 3.1             & 14              & 2.0$''$$\times$1.6$''$\phantom{0} ($-$76$^\circ$) \\ 
L1527  & 6.6             & 4.2             & 1.6$''$$\times$0.61$''$ ($+$79$^\circ$)           & 3.3             & 4.5             & 1.8$''$$\times$1.2$''$\phantom{0} ($-$70$^\circ$) \\ 
L483   & 5.8             & 5.5             & 2.1$''$$\times$1.6$''$\phantom{0} ($+$86$^\circ$) & 3.0             & 6.4             & 2.4$''$$\times$2.1$''$\phantom{0} ($+$70$^\circ$) \\ 
B335   & 5.1             & 11              & 2.1$''$$\times$1.8$''$\phantom{0} ($+$82$^\circ$) & 2.8             & 10              & 2.5$''$$\times$2.1$''$\phantom{0} ($+$46$^\circ$) \\ 
L1157  & 6.8             & 9.5             & 2.1$''$$\times$1.6$''$\phantom{0} ($-$30$^\circ$) & 4.2             & 9.4             & 2.6$''$$\times$2.2$''$\phantom{0} ($-$36$^\circ$) \\ \hline
\end{tabular}
\tablenotetext{a}{Theoretical noise (RMS).} \tablenotetext{b}{Cleaned map noise (RMS).} \tablenotetext{c}{Size and position angle.}
\end{table}
\clearpage
\section{Results: Continuum emission}\label{contresults}
\subsection{Morphology}\label{cont_morph}
\begin{figure}
\resizebox{\hsize}{!}{\includegraphics{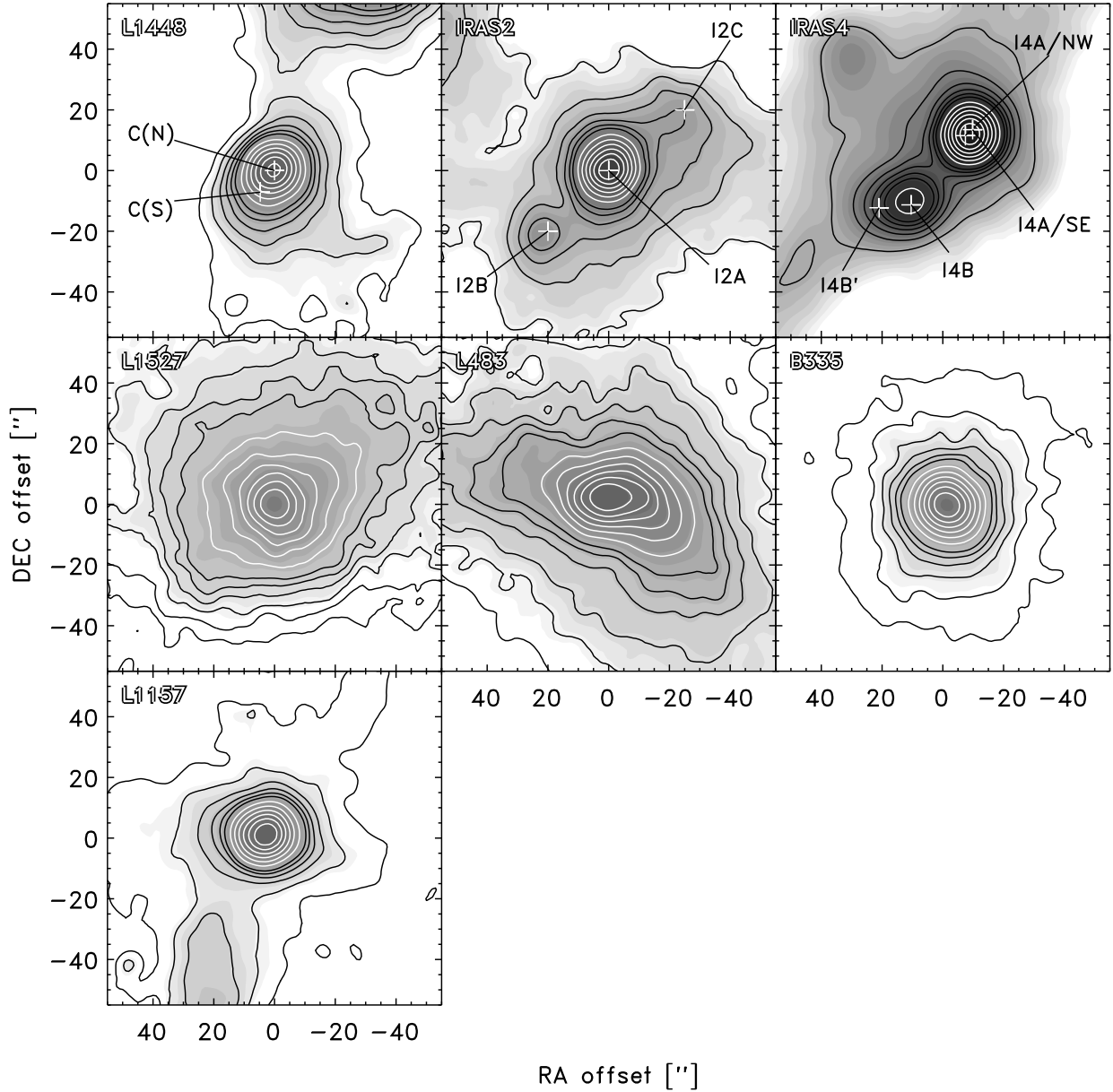}}
\caption{SCUBA maps of the eight sources in our sample from the JCMT
  archive. In each map the coordinates are given relative to the
  brightest emission, except in the IRAS4 where a point between the
  two sources has been chosen as reference. The contour levels are
  given relative to the peak flux at 5\%, 10\%, 15\%, $\ldots$, 30\%
  (black contours) and from there at 40\%, 50\%, 60\%, $\ldots$, 90\%
  (white contours). In the L1448, IRAS2 and IRAS4 maps, the white plus
  symbols and labels indicate the individual sources discussed in the
  text. For further details about the JCMT/SCUBA maps of these
  regions, see, e.g., \cite{chandler00}, \cite{shirley00}, and
  \cite{sandell01}.}\label{scuba_continuum}
\end{figure}

\begin{figure}
\vspace*{-15mm}
\resizebox{0.6\hsize}{!}{\includegraphics{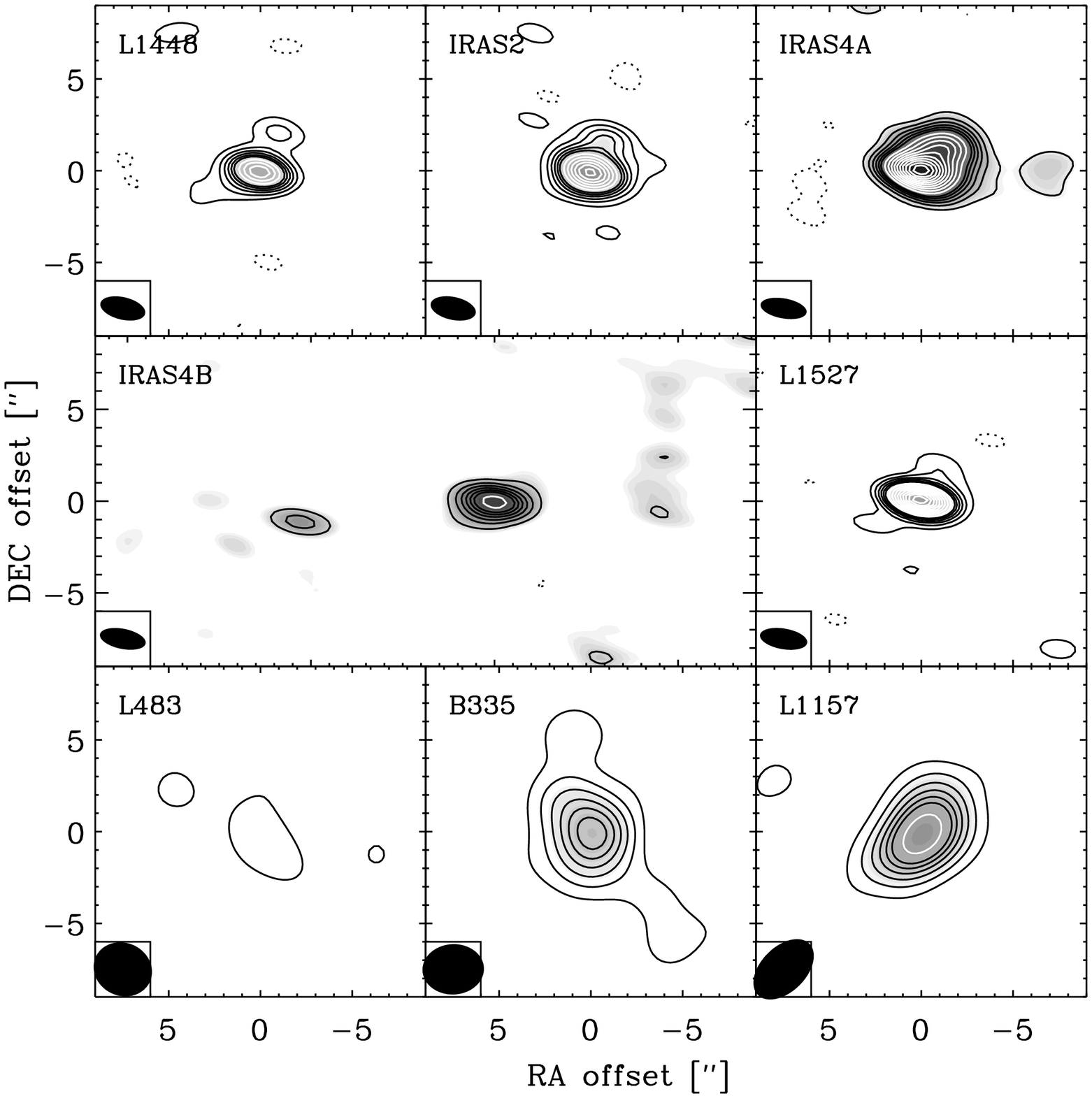}}

\resizebox{0.6\hsize}{!}{\includegraphics{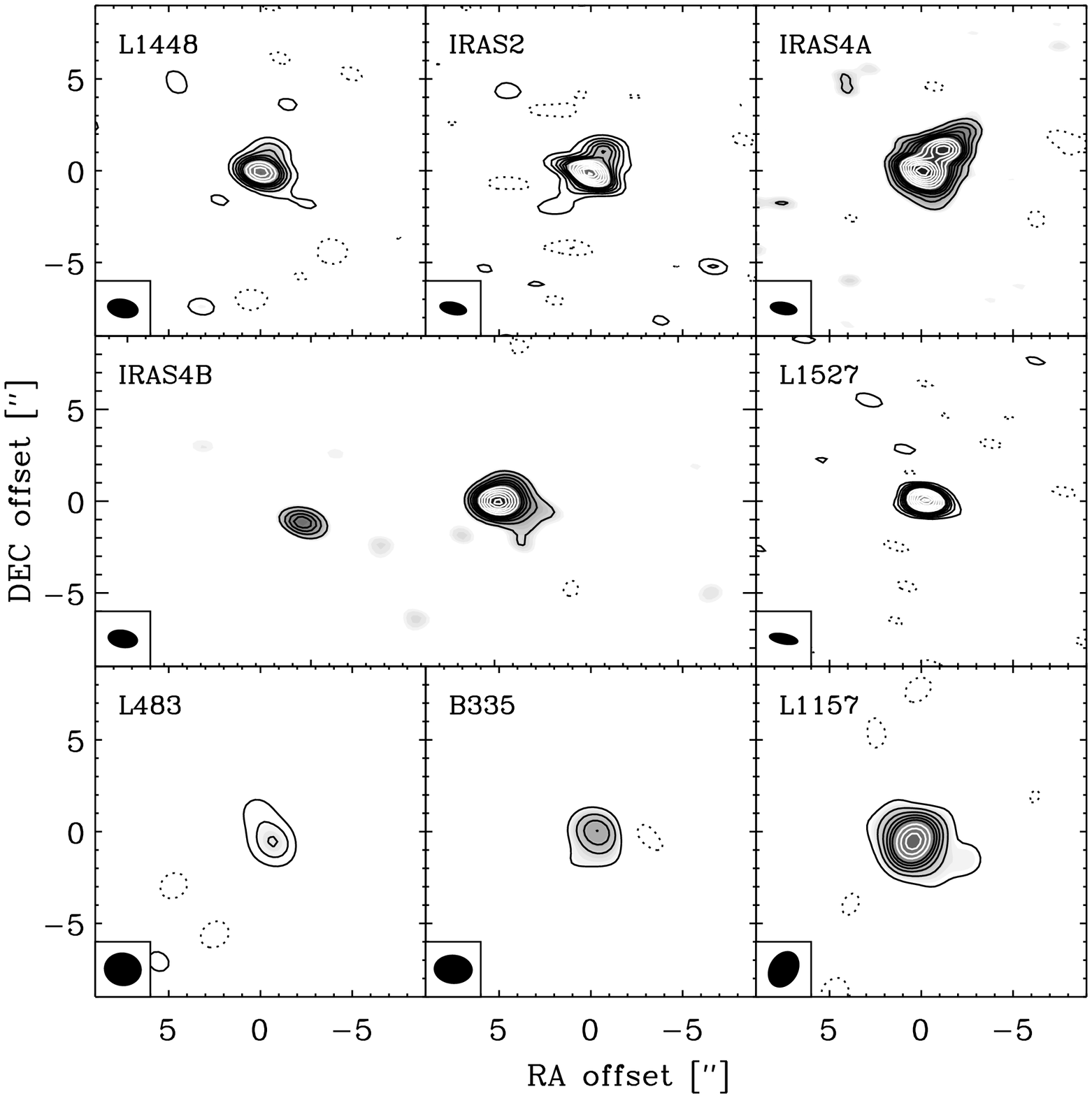}}
\caption{Maps of each of the fields from the 1.3~mm observations
  (upper panels) and 0.8~mm observations (lower panels) restored with
  uniform weighting. In each panel, the black contours are given in
  steps of 3$\sigma$ up to 18$\sigma$ followed by white contours in
  steps of 6$\sigma$ at 24$\sigma$, 30$\sigma$, 36$\sigma$,
  $\ldots$. Note that the cores in the lowest row of panels (L483,
  B335 and L1157) were observed in the compact configuration which
  gives slightly better sensitivity to extended emission, but lower
  resolution than the remaining sources observed in the Compact-North
  configuration.}\label{continuum_maps}
\end{figure}

Figure~\ref{scuba_continuum} shows SCUBA maps from the JCMT
archive\footnote{The JCMT archive at the Canadian Astronomy Data
  Centre is operated by the Herzberg Institute of Astrophysics,
  National Research Council of Canada.} and
Figure~\ref{continuum_maps} continuum maps for each pointing from the
SMA observations. The SCUBA cores extend over more than 50\arcsec\
($\approx 10,000$~AU) in almost all cases, so the interferometer
clearly resolves out a very significant fraction of the extended
emission (i.e., the protostellar envelopes) and only zooms in on the
compact emission in the center. Still, comparing to the same sources
observed, e.g., in $^{13}$CO or C$^{18}$O 2--1 (see following
section), it is noteworthy that the difference between the continuum
signatures of the sources observed in the Compact and Compact-North
configurations respectively are less than in the CO isotopic lines. It
suggests that the continuum emission is dominated by a central compact
component not related to the envelope, but more likely associated with
the central circumstellar disk. These central components are most
easily identified in the visibility curves at the longest baselines,
where the flux should approach a non-zero constant value. The SCUBA
maps of L1527 and L483 appear different than those of the other
sources in the sample. Figure~\ref{scuba_continuum} clearly
illustrates that the larger scale emission from these cores is
different morphologically than the remaining sources with a larger
degree of extended emission, possibly reflecting flattened brightness
profiles. For L483 in particular it is found that most of the
continuum emission in the SMA data is related to this more extended
core with little compact emission seen on the longest baselines.

A number of the sources are double or have fainter companions within
the SMA primary beam field of view
(Fig.~\ref{scuba_continuum}). L1448-C has been resolved into two
components at mid-infrared wavelengths by recent Spitzer observations
\citep{perspitz}, L1448-C(N) and L1448-C(S) with a separation of
8$''$, the northern source L1448-C(N) being the well-known source from
millimeter interferometric studies, L1448-mm. Faint continuum emission
is seen at the 3--5$\sigma$ levels at 1.3~mm and 0.8~mm toward the
southern source of L1448-C although the northern source is clearly
dominating the maps. IRAS2A also has two accompanying cores (IRAS2B
and IRAS2C) identified from SCUBA maps \citep{sandell01} and high
angular resolution continuum and line studies
\citep[e.g.,][]{looney00,n1333i2art}. Of these, IRAS2B (at 30$''$
distance) is found to fall outside the primary beam field of view at
0.8~mm but inside at 1.3~mm where it clearly is detected. IRAS2C is
likely a starless core \citep{n1333i2art} and not identified in these
continuum maps. The companion to IRAS4B, IRAS4B$'$, is clearly
detected at both wavelengths at a separation of about 11$''$. IRAS4A
is resolved into its two components (separation of 2$''$) at both
wavelengths.

For a number of the remaining sources, faint extended continuum
emission is seen. The continuum emission toward the central source
L1448-C(N) appears to be slightly extended to the northeast and break
up into two at a separation of about 1--2$''$. Likewise, the 0.8~mm
observations of IRAS2A appear to resolve the source with extended
emission toward the North. It is likely that IRAS2A itself is a binary
given the quadrupolar morphology of its outflow on small scales
\citep[see discussion in][]{n1333i2art}. Still, given that a
significant fraction of the envelope emission is resolved out at the
central position, it is not obvious whether these additional features
in the extended emission really represent nearby companions or whether
they simply reflect the particular $(u,v)$ coverage.

\clearpage

\thispagestyle{empty}
\clearpage
\subsection{Dust continuum emission from protostellar disks}
Table~\ref{continuum_gauss} and \ref{continuum_point} list the derived
parameters for each of the bright components from circular Gaussian
fits to the visibilities and from point source fits to the longest
baselines ($\ge$~40~k$\lambda$), respectively. The statistical
uncertainties on the derived parameters are generally small: the
uncertainties on the peak and total fluxes are the largely determined
by the RMS noise levels of the data (Table~\ref{rms}) typically
3--5~mJy; somewhat larger, $\approx$10~mJy, for the bright binary
sources NGC~1333-IRAS4A and -IRAS4B. On the other hand the
calibrational uncertainty on these fluxes is typically 20--30\%,
therefore dominating the uncertainties on the derived fluxes. Likewise
the statistical uncertainties on the positions and sizes of the
Gaussians are $\sim 0.1''$ or better - largely determined by the S/N
of the detections. However, these statistical uncertainties rely on,
e.g., the circular Gaussians representing the true distribution of the
emission which is likely not to be the case.  Therefore, the actual
uncertainties on these parameters are larger - but difficult to
quantify without more specific models for the distribution of the
observed emission.

Comparisons of the total fluxes obtained from the Gaussian fits to the
SCUBA maps at 850~$\mu$m (Fig.~\ref{scuba_continuum};
\citealt{chandler00}, \citealt{shirley00} and \citealt{sandell01})
suggest that typically only 10--20\% of the extended emission seen
with SCUBA is recovered by the SMA observations. This is also
suggested from a comparison between the Gaussian fits at the two
frequencies observed with the SMA. Typically the sizes of the
Gaussians fit to the 1.3~mm continuum emission are found to be larger
than those at 0.8~mm consistent with these sizes reflecting the
slightly shorter baselines (in units of k$\lambda$) for the longer
wavelength data and thus less resolving out. Also, for a number of the
sources, the positions derived for the 230~GHz and 345~GHz continuum
observations differ by up to 1\arcsec, which is unlikely due to actual
differences in the positions but rather reflect the uneven $(u,v)$
sampling. Furthermore, for the sources with the most centrally peaked
emission, the best agreement is found between the positions suggesting
that the emission from these sources is more weighted to the compact
(disk) emission at the center and the envelope mostly resolved
out. Fitting the continuum emission with point sources at the longest
baselines gives slightly better agreements between the positions from
the 0.8~mm and 1.3~mm continuum data. It should be emphasized that
some of the sources show variations in fluxes also at longer baselines
indicating that the central component is resolved \citep[e.g.,
IRAS2A][]{iras2sma} and for those the fitted point source flux at the
given wavelength is an underestimate of the total flux of the compact
component.

\begin{figure}
\resizebox{\hsize}{!}{\includegraphics{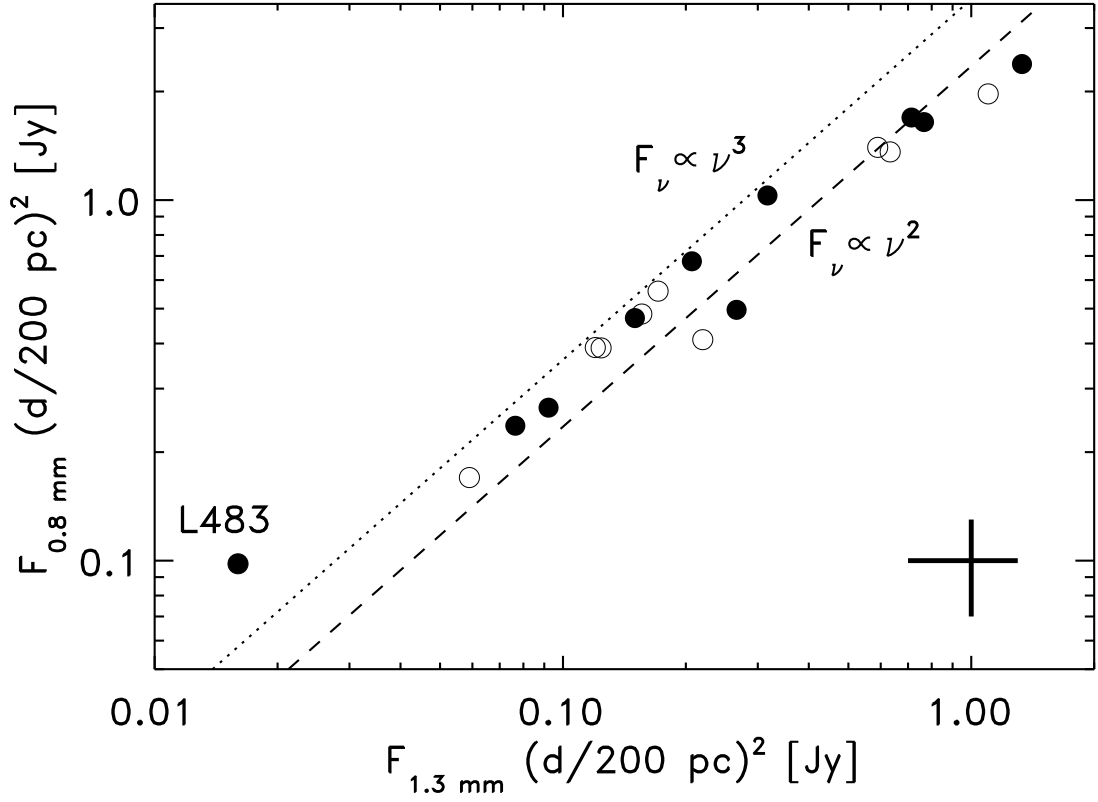}}
\caption{Comparison of the point source fluxes obtained from fits to
  baselines longer than 40~k$\lambda$. The open symbols show the
  actual derived fluxes whereas the filled symbols show the fluxes
  normalized to a distance of 200~pc. The typical calibrational error
  of $\pm 30$\% is indicated in the lower
  right.}\label{flux_comparison}
\end{figure}

Figure~\ref{flux_comparison} shows a comparison between the 0.8~mm and
1.3~mm continuum fluxes from the point source fits. The point source
continuum fluxes at 0.8~mm and 1.3~mm correlate nicely, as expected if
they are probing the same unresolved component. It is noteworthy that
the continuum flux varies by more than an order of magnitude between
the sources at either wavelengths. The only outlier, which does not
show a strong detection at the longer baselines, is L483 consistent
with the 3~mm observations by \cite{l483art}.

Interestingly all other sources show compact emission on the longer
baselines, which in some cases have been attributed to contributions
of central circumstellar disks through models of the observed
visibilities \citep[e.g.,][]{keene90,hogerheijde99,looney03}. One
problem about such models is the degeneracy between the contributions
of disk and envelope emission on small scales and in general it is
important to use constraints on the envelope physical structures (such
as temperature, density and mass) from other independent
observations. A particular powerful method has turned out to be
detailed dust radiatve transfer models like those presented for the
sources in this sample \citep{jorgensen02}. Such models can be used to
predict the emission observed by the interferometric observations and
address how much the large scale envelope emits on the long
baselines/small scales probed by the interferometer and whether an
additional component, such as a disk, exists besides this
envelope. Four out of the eight sources in this sample have been
analysed in this way, specifically: NGC1333-IRAS2A:
\citep{n1333i2art,iras2sma}; L1448-mm: \citep{hotcorepaper}; L483:
\citep{l483art}; B335: \citep{harvey03}. These studies typically find
that less than $\approx$10--30\% of the continuum emission at
baselines longer than 40~k$\lambda$ can be attributed to the extended
envelope component (see, e.g., Fig. 3 of \citealt{iras2sma}). It is
therefore as a first approximation reasonable to attribute most of the
emission on these longer baselines to a component separate from the
envelope.

The strong correlation provides important clues to the nature of these
central unresolved sources. For optically thin emission, the slope
between the two frequencies, $\alpha$, will be related to the slope,
$\beta$, of the dust opacity law, $\kappa \propto \nu^\beta$, as
$\alpha \approx \beta + 2$ in the Rayleigh-Jeans limit
\citep{beckwith91}. Figure~\ref{flux_comparison} shows that the
spectral slopes for all sources are lower than 3.0 with a few lower
than 2.0 (average value of 2.4; median 2.6). The optically thin
envelope emission would have a spectral slope of 3.5--4 for typical
dust opacities which are successfully used to explain the larger scale
envelopes in detailed dust radiative transfer models (for example
$\beta \approx 1.8$ at 850~$\mu$m for the coagulated dust grains, the
so-called ``OH5 dust'', of \cite{ossenkopf94} used in
\cite{jorgensen02}). The protostellar envelopes are not optically
thick at 850~$\mu$m - or we should not detect any mid-infrared
emission from the central protostars; e.g., \cite{perspitz}. Slopes of
3.5--4 are also what were observed by \cite{dent98} in a single-dish
survey of the continuum emission from the envelopes around embedded
YSOs. We note that since the envelope emission is expected to have a
spectral index of 3.5--4, any contribution to the derived point source
fluxes will therefore contribute more to the emission at 345 GHz
relative to that at 230 GHz compared to the observed flatter spectral
indices of 2--3. Thereby subtracting any envelope contribution will in
fact flatten the derived spectral index of the compact component.

The flatter spectral indices, $\alpha < 3.0$ seen here, suggest that
the compact continuum emission has its origin in a different
component, likely the disk. Furthermore the spectral indices suggest
the compact disk emission is optically thick, and some of the
variations in disk fluxes more likely reflect differences in disk
sizes and temperatures, rather than mass. For a disk of mass
(gas+dust), $M_D$ and radius $R_D$ the average optical depth is
\citep{beckwith91}:
\begin{equation}
\langle\tau_\nu\rangle = \frac{\kappa_\nu/100 M_D}{\pi R_D^2 \cos \theta}
\end{equation}
where $\theta$ is the disk inclination angle and $\kappa_\nu$ is the
opacity per dust mass (e.g., 1.75~cm$^2$~g$^{-1}$ at 870~$\mu$m for
OH5 dust), which for typical disk parameters (total gas+dust mass of
0.1~$M_\odot$ and a radius of 100~AU) can be expressed as:
\begin{equation}
\langle\tau_{870}\rangle = 0.5/\cos\theta \left(\frac{M_D}{0.1 M_\odot}\right)\left(\frac{R_D}{100 {\rm AU}}\right)^{-2}
\end{equation}
For example, a lower limit to the disk mass of 0.3~$M_\odot$ was
derived for NGC1333-IRAS2A from 3~mm observations by \cite{n1333i2art}
whereas the disk size was estimated to be 150~AU (radius) in
\cite{iras2sma}. This would imply a lower limit to the average optical
thickness of about 0.7 at 850~$\mu$m and 0.3 at 1.3~mm (assuming the
disk is seen face-on). In terms of its flux, IRAS2A appears typical
for the sample of the objects, suggesting that the emission from all
the disks in our sample is marginally optically thick. That the
emission is marginally optically thick could in fact explain the
variations in the derived spectral indices: if any of the highest
derived spectral indices had their origin in optically thin emission,
the constraint $\alpha < 3$ would imply $\beta \lesssim 1$, lower than
the $\beta = 2$ for typical dust in the interstellar medium. The
values of $\alpha \approx 2.0$ would then represent the disks being
optically thick. For more evolved T-Tauri and Herbig Ae/Be stars,
grain growth has been inferred as an explanation for values of $\beta$
in the range of 0.5--1.0 \cite[e.g.,][]{mannings94,
  natta04,andrews05,draine06,rodmann06,lommen06}. \cite{draine06}
shows that for ISM type dust, $\beta \approx 1$ at 1~mm if the size
distribution of the dust grains ($dn/da \propto a^{-p}$ with $p\approx
3.5$) extends to sizes $a_{\rm max} \gtrsim 3$~mm. On the other hand,
if the spectral indices are found to be about $2$ all the way to cm
wavelengths such as seen in a few objects (IRAS2A, \citealt{iras2sma};
L1448-C, \citealt{hotcorepaper}; IRAS16293-2422B
\citealt{hotcorepaper,chandler05}), this would imply even larger
maximum grain sizes.

If the disks are marginally optically thick, the observed continuum
fluxes provide a lower limit to their masses. For optically thin
emission from a disk with a single temperature, the disk mass is:
\begin{equation}
M=\frac{S_\nu d^2}{\kappa_\nu B_\nu(T)}
\end{equation}
or for typical parameters of the dust temperature (30~K) and opacities
at 1.3 and 0.8~$\mu$m \citep{ossenkopf94}:
\begin{eqnarray}
M_{\rm 1.3 mm} & = & 1.3\,M_\odot \left(\frac{F_{\rm 1.3~mm}}{\rm 1~Jy}\right)\left(\frac{d}{\rm 200~pc}\right)^2\left(\exp\left[0.36\left(\frac{\rm 30~K}{T}\right)\right]-1\right) \\
M_{\rm 0.8 mm} & = & 0.18\,M_\odot \left(\frac{F_{\rm 0.8~mm}}{\rm 1~Jy}\right)\left(\frac{d}{\rm 200~pc}\right)^2\left(\exp\left[0.55\left(\frac{\rm 30~K}{T}\right)\right]-1\right)
\end{eqnarray}
Using these parameters, Table~\ref{continuum_point} lists the total
masses (gas+dust) of the disks. Given that these disks are likely
optically thick rather than thin, the derived masses are lower
limits. These disks appear to be at least as massive as typical
T-Tauri disks \citep{andrews05} (or perhaps more directly; they have
comparable fluxes), arguing against a significant build up of the
disks from the Class 0 stage to the later T-Tauri stages. Typically
the inferred disk masses constitute about 1--10\% of the masses of the
circumstellar envelopes derived by \cite{jorgensen02} (listed in
Table~\ref{continuum_point}) and are non-negligible reservoirs of
mass. This fraction is also similar to what was found by
\cite{looney03} who quote compact sources present in their sample of
Class 0 sources of 0--0.12~$M_\odot$ compared to envelope masses of
$\approx 1$~$M_\odot$. We emphasize again that these mass estimates
are uncertain due to the adopted dust opacities by a factor 2--3 in
itself, with additional uncertainties introduced by other parameters
such as the disk temperature and exact contribution from the
envelope. A detailed comparison to the envelope models and to models
for the potential disk structures is required to address these issue,
but is outside the scope of this paper.

In summary, the compact continuum emission seen toward the observed
protostellar sources likely comes from the circumstellar disks with
circumstellar envelopes contributing somewhat at the very shortest
baselines. The large number of sources studied homogeneously in this
sample allow for direct comparisons between the continuum fluxes and
inferred properties of the circumstellar disks. The spectral slopes
inferred from the 0.8~mm and 1.3~mm data together with previous longer
wavelength observations of a few sources suggest that the emission is
marginally optically thick and that the slope of the dust opacity law
is $\beta \approx 1$, indicative of grain growth already in the young
disks around deeply embedded protostars. The disks are found to
contain a significant amount of material with a typical lower limit to
the total (gas+dust) mass of about 0.1~$M_\odot$ (dependent on
assumptions about dust temperatures, opacities etc.), i.e., about
1--10\% of the envelope masses. Compared to more evolved T-Tauri
stars, the properties of these disks argue against a significant
build-up of disks after the deeply embedded stages - or as suggested
by \cite{looney03}, that the processing of material through the disks
in the Class 0 stage occur rapidly to avoid a significant build-up of
mass. An important aspect of further studies will be to address fully
the relative contribution to the continuum emission from the envelope
and disks and to model the radial variations of surface densities and
optical depth in disks for the sources for which the central continuum
component is resolved (e.g., IRAS2A, \cite{iras2sma}).
\begin{table}
  \caption{Results of circular Gaussian fits to the continuum visibilities.}\label{continuum_gauss}
\begin{tabular}{llllllll} \tableline\tableline\small
Source & \multicolumn{3}{c}{Position} & \multicolumn{2}{c}{Flux density} & \multicolumn{2}{c}{Size (FWHM)} \\
       & $\alpha$ (0.8 mm) & $\delta$ (0.8~mm) & Offset (1.3~mm) & $F_{\rm 1.3~mm}$ & $F_{\rm 0.8~mm}$ & $\theta_{1.3 {\rm~mm}}$ & $\theta_{0.8 {\rm~mm}}$ \\
       & (J2000.0)         & (J2000.0)         & (arcsec)        & [Jy]        & [Jy]        &   [$''$]                 &  [$''$]                   \\ \tableline
L1448     &  03 25 38.87 & $+$30 44 05.4 & ($+$0.10 ,$+$0.02) & 0.18  & 0.53 & 0.94 & 0.77 \\
IRAS2A     &  03 28 55.58 & $+$31 14 37.1 & ($+$0.00 ,$+$0.08) & 0.34  & 0.90 & 1.4  & 0.94 \\
IRAS4A-SE &  03 29 10.54 & $+$31 13 30.9 & ($+$0.08 ,$+$0.09) & 1.8   & 3.5  & 1.2\tablenotemark{a} & 1.2\tablenotemark{a} \\
IRAS4A-NW &  03 29 10.44 & $+$31 13 32.2 & ($-$0.03 ,$+$0.09) & 1.0   & 2.4  & 1.2\tablenotemark{a} & 1.2\tablenotemark{a} \\
IRAS4B    &  03 29 12.01 & $+$31 13 08.1 & ($+$0.24 ,$-$0.04) & 0.92  & 2.1  & 1.0  & 1.0  \\
IRAS4B$'$ &  03 29 12.83 & $+$31 13 06.9 & ($+$0.28 ,$-$0.01) & 0.25  & 0.47 & 0.68 & 0.63 \\
L1527     &  04 39 53.88 & $+$26 03 09.8 & ($+$0.26 ,$-$0.01) & 0.19  & 0.58 & 0.69 & 0.75 \\
L483      &  18 17 29.92 & $-$04 39 39.5 & ($+$0.60 ,$-$0.52) & 0.044 & 0.20 & 3.7  & 1.7  \\
B335      &  19 37 00.91 & $+$07 34 09.6 & ($+$0.42 ,$+$0.18) & 0.16  & 0.35 & 2.9  & 1.6  \\
L1157     &  20 39 06.28 & $+$68 02 15.8 & ($-$0.42 ,$+$0.47) & 0.30  & 0.76 & 2.6  & 1.7  \\ \tableline
\end{tabular}
\tablecomments{Units of right ascension are hours, minutes, and seconds and units of declination are degrees, arcminutes and arcseconds. See the text for a discussion of the uncertainties of the derived parameters.} 
\tablenotetext{a}{For the fits to the IRAS4A continuum data a Gaussian FWHM of 1.2$''$ was assumed (and fixed).}
\end{table}

\begin{table}
  \caption{Results of point source fits to the continuum visibilities at
    baselines longer than 40~k$\lambda$ and mass estimates assuming optical thin dust emission.}\label{continuum_point}
\begin{tabular}{llllllll} \tableline\tableline\small
Source & \multicolumn{3}{c}{Position} & \multicolumn{2}{c}{Flux density} \\
       & $\alpha$ (0.8 mm) & $\delta$ (0.8~mm) & Offset (1.3~mm) & $F_{\rm 1.3 mm}$ & $F_{\rm 0.8 mm}$  & $M_{\rm 30 K}$\tablenotemark{a} & $M_{\rm env}$\tablenotemark{b} \\
       & (J2000.0)         & (J2000.0)  & (arcsec) & [Jy]        & [Jy]           & [$M_\odot$]      & [$M_\odot$] \\ \tableline
L1448\tablenotemark{c} &  03 25 38.87  & $+$30 44 05.4 & ($+$0.13 ,$-$0.12) & 0.12   & 0.37  &  0.082 / 0.060   & 0.93 \\
IRAS2A      &  03 28 55.58  & $+$31 14 37.1 & ($+$0.05 ,$+$0.01) & 0.17   & 0.48  &  0.12 / 0.078     & 1.7 \\
IRAS4A-SE  &  03 29 10.53  & $+$31 13 31.0 & ($+$0.11 ,$+$0.01) & 1.1    & 2.0   &  0.76 / 0.32      & 2.3 \\
IRAS4A-NW  &  03 29 10.44  & $+$31 13 32.3 & ($+$0.18 ,$-$0.11) & 0.59   & 1.4   &  0.41 / 0.23      & -- \\
IRAS4B     &  03 29 12.01  & $+$31 13 08.1 & ($+$0.25 ,$-$0.11) & 0.59   & 1.4   &  0.41 / 0.23      & 2.0 \\
IRAS4B$'$  &  03 29 12.83  & $+$31 13 07.0 & ($+$0.18, $-$0.08) & 0.22   & 0.40  &  0.15 / 0.065     & -- \\
L1527      &  04 39 53.88  & $+$26 03 09.8 & ($+$0.27 ,$+$0.51) & 0.15   & 0.38  &  0.072 / 0.025    & 0.91 \\
L483       &  18 17 29.91  & $-$04 39 39.6 & ($-$0.18 ,$-$0.48) & 0.016  & 0.097 &  0.009 / 0.012    & 4.4 \\
B335       &  19 37 00.91  & $+$07 34 09.7 & ($-$0.13 ,$+$0.04) & 0.059  & 0.17  &  0.052 / 0.035    & 2.6 \\
L1157      &  20 39 06.28  & $+$68 02 16.0 & ($-$0.38 ,$-$0.01) & 0.12   & 0.39  &  0.18 / 0.14      & 1.6 \\ \tableline
\end{tabular}

\tablecomments{Units of right ascension are hours, minutes, and seconds and units of declination are degrees, arcminutes and arcseconds. See the text for a discussion of the uncertainties of the derived parameters.}
\tablenotetext{a}{Mass derived from 1.3~mm and 0.8~mm continuum data, respectively.}
\tablenotetext{b}{Mass of the envelope from the dust radiative transfer models of \cite{jorgensen02} with the exception of B335 for which the estimate from \cite{shirley02} is adopted. No (separate) values are given for IRAS4A-NW and IRAS4B$'$ as these are likely included in the circumbinary systems together with IRAS4A-SE and IRAS4B, respectively.}
\tablenotetext{c}{Refers to the Northern component, L1448-C(N)
\citep{perspitz}. The Southern source, L1448-C(S) at (03:25:39.14;
+30:43:58.3), is detected with point source fluxes of 12.8~mJy (230
GHz; RMS $\approx 3$~mJy) and 43.5~mJy (345~GHz; RMS $\approx 8$~mJy).}
\end{table}

\clearpage
\section{Results: Line emission}\label{lineresults}
\begin{figure}
\resizebox{\hsize}{!}{\includegraphics{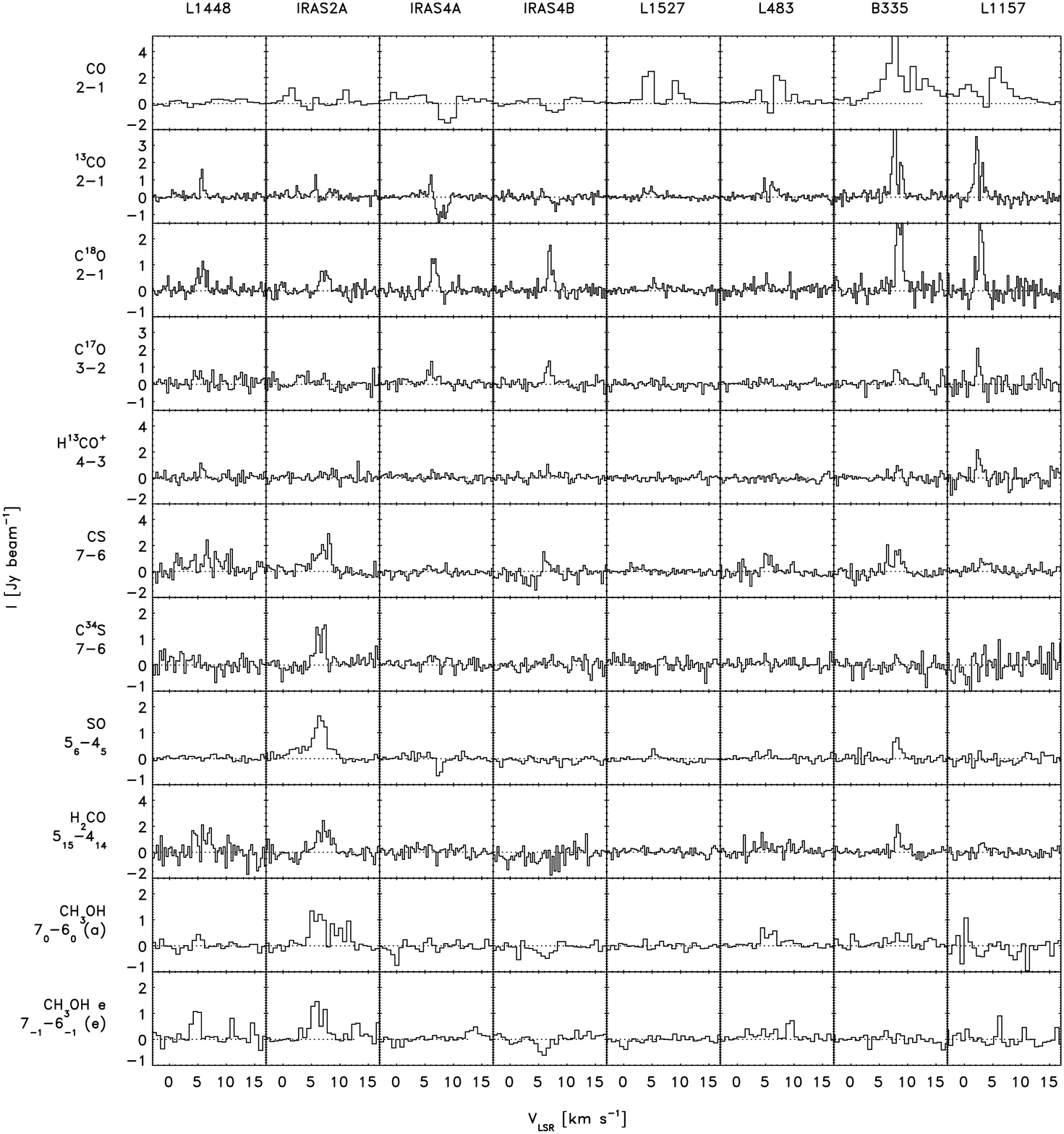}}
\caption{Spectra toward the center position of each source.}\label{firstspectrum}
\end{figure}

\begin{figure}
\resizebox{!}{0.9\vsize}{\includegraphics{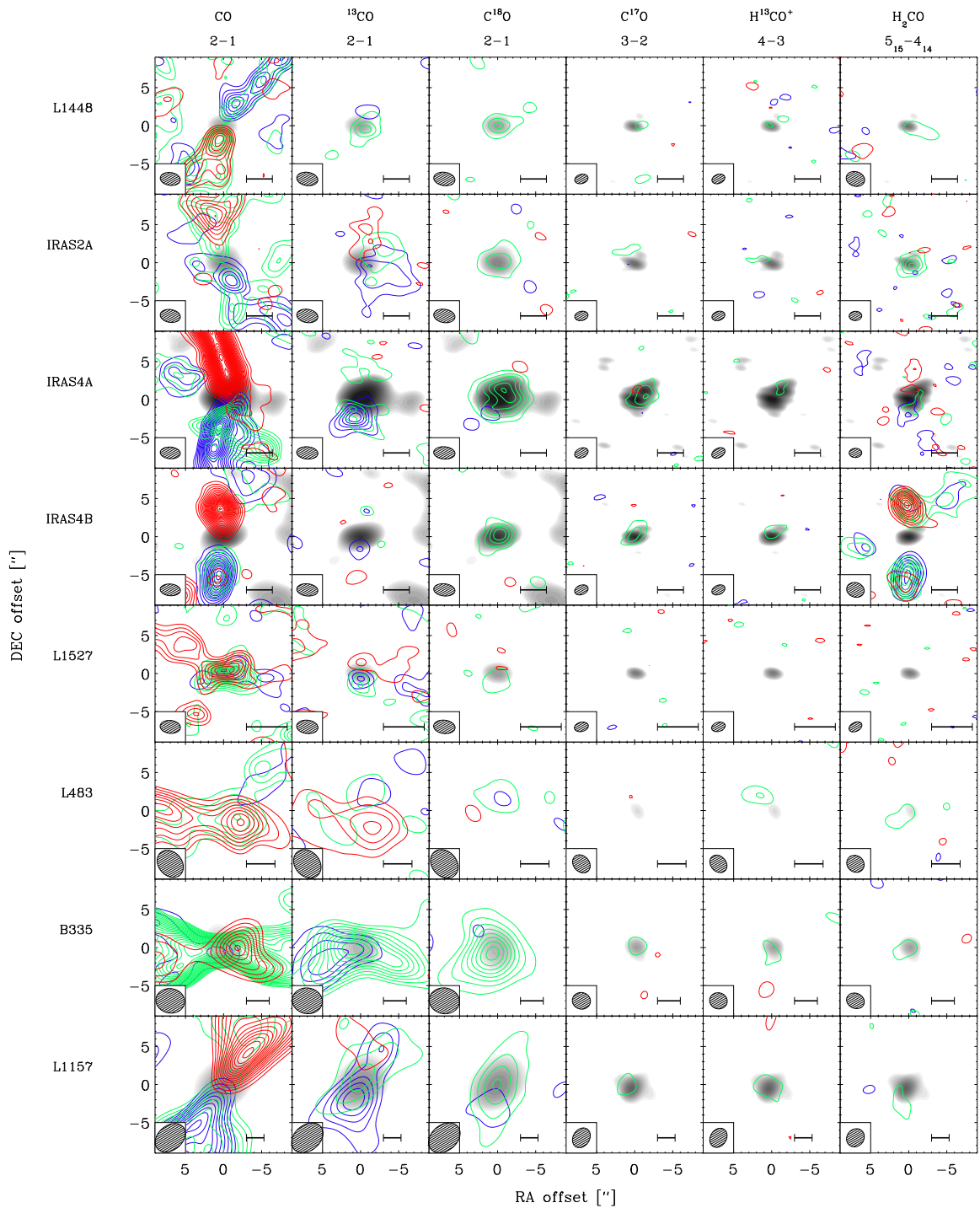}}
\caption{Overview of the entire set of SMA data from the
  program. Contours given in steps of 3$\sigma$ integrated over the
  velocity intervals specified in the text (green/blue/red indicate
  material at the systemic velocity and blue/red shifted from that,
  respectively). The scale bar in the lower right corner of each panel
  represents a size of 750~AU toward each source, respectively. The
  grey scale indicates the continuum emission at 0.8~mm and 1.3~mm
  (corresponding to the shown line) from
  Figure~\ref{continuum_maps}.}\label{firstmolmap}
\end{figure}

Figure~\ref{firstspectrum} shows the spectra in the central beam
toward the continuum peak positions for each detected line while
Figure~\ref{firstmolmap} shows maps of the integrated line intensities
over three velocity intervals: $[V_{c}-1.5\Delta V; V_{c}-0.5\Delta
V]$, $[V_{c}-0.5\Delta V; V_{c}+0.5\Delta V]$ and $[V_{c}+0.5\Delta V;
V_{c}+1.5\Delta V]$ around the system velocity $V_c$ for each
source. $\Delta V$ and $V_c$ were chosen based on the widths of
Gaussian fitted to the C$^{18}$O line at the center positions
(Table~\ref{gaussfit_lines}) with $\Delta V=a\,\sigma$ where $\sigma$
is the Gaussian velocity dispersion ($\sigma={\rm
  FWHM}/(2\sqrt{2\ln(2)}$) and $a=1.5$ for $^{13}$CO, C$^{18}$O 2--1,
C$^{17}$O 3--2, H$^{13}$CO$^+$ 4--3 and C$^{34}$S 7--6, $a=3$ for SO
$5_6-4_5$, H$_2$CO $5_{15}-4_{14}$ and CS 7--6, and $a=6$ for CO 2--1,
CH$_3$OH $7_k-6_k$ and SiO 7--6. Table~\ref{linedetections} lists the
lines detected toward each source.

\begin{table}
\caption{Result of Gaussian fits to C$^{18}$O 2--1 line profiles at the central positions of each source.}\label{gaussfit_lines}
\begin{tabular}{lll}\hline\hline
Source & $V_c$$^{a}$ [km~s$^{-1}$] & FWHM$^{b}$ [km~s$^{1}$] \\ \hline
L1448  & 5.7 & 4.3 \\
IRAS2A  & 7.3 & 4.3 \\
IRAS4A & 6.6 & 2.7 \\
IRAS4B & 7.0 & 2.0 \\
L1527  & 5.4 & 2.0 \\
L483   & 5.2 & 2.0 \\
B335   & 8.5 & 3.0 \\
L1157  & 2.9 & 2.2 \\ \hline
\end{tabular}

$^{a}$Systemic velocity: statistical uncertainties from Gaussian fits of 0.05--0.1~\kms. $^{b}$Line width (FWHM): statistical uncertainties from Gaussian fits of 0.1--0.3~\kms.
\end{table}

All sources show detections in the 2--1 transition of the CO isotopic
species at 230~GHz and most sources are detected in the expected
strong lines of the main isotopologues of CS 7--6 and H$_2$CO
$5_{15}-4_{14}$ lines at 342.8 and 351.7~GHz, with L1527 being the
exception. For the other lines IRAS2A shows strong centrally condensed
emission close to the central protostar in all species, whereas the
other sources in the sample have detections in only some of the lines
with extended emission in some cases.

\begin{deluxetable}{lcccccccc}
  \tablecaption{Detected molecular lines.\label{linedetections}}

\tablehead{\colhead{Line} & \colhead{L1448} & \colhead{IRAS2A} & \colhead{IRAS4A} & \colhead{IRAS4B} & \colhead{L1527} & \colhead{L483} & \colhead{B335} & \colhead{L1157}}
\startdata
\multicolumn{9}{c}{230 GHz setting}\\ \tableline  
$^{12}$CO 2--1                         & + & + & + & + & + & + & + & + \\
$^{13}$CO 2--1                         & + & + & + & + & + & + & + & + \\
SO  $5_6-4_5$                          &(O)& + & + & + & + &+* & + &(O)\\ 
C$^{18}$O 2--1                         & + & + & + & + & + & + & + & + \\
CH$_3$OH                               & - & - & - &(O)& - & - & - & - \\ \tableline
\multicolumn{9}{c}{337 GHz setting} \\ \tableline  
C$^{17}$O 3--2                         &(+)&(+)& + & + & - & - & + & + \\
C$^{34}$S 7--6                         & - & + &(+)& - & - & - & - & - \\
CH$_3$OH $7_0-6_0$\tablenotemark{a}  &(+)\tablenotemark{d} & + & O\tablenotemark{e} & O & - & + & - & - \\
CH$_3$OH $7_{-1}-6_{-1}$\tablenotemark{b}  &(+)& + & O & O & - & + & - & - \\
CH$_3$OH $7_{\pm 2}-6_{\pm 2}$\tablenotemark{b}     & - & + & O & O & - & - & - & - \\ 
CH$_3$OH $7_k-6_k$\tablenotemark{c}       & - & + & - & - & - & - & - & - \\
SiO 8--7                               & - & - &(O)& O & - & - & - & - \\
HN$^{13}$C 4--3                        & - & - & - & - & - & - & - & - \\
H$^{13}$CO$^+$ 4--3                    & + & - & - &+* & - &+* & - & + \\ \tableline
\multicolumn{9}{c}{342 GHz setting:} \\ \tableline 
CS 7--6                                & + & + &+/O&+/O& - & + & + & + \\
H$_2$CO $5_{15}-4_{14}$                & + & + &+/O& O & - &(+)& + & + \\
SO$_2$                                 &(+)& - & - & - & - & - & - & - \\ \tableline
\enddata
\tablecomments{``+'' indicates that line emission detected above 3$\sigma$ toward the
central source position, ``O'' that the line is detected in the
outflow offset from the central protostar, ``*'' that line emission is
detected offset from the central protostar and also not obviously
connected to any outflow shocks and ``-'' that no emission is
detected. Symbols given in parenthesis indicate tentative detections.}
\tablenotetext{a}{A-type CH$_3$OH.}  \tablenotetext{b}{E-type
  CH$_3$OH.}  \tablenotetext{c}{Multiple high excitation transitions
  observable; see Figure~\ref{ch3oh_spectra}.}
\tablenotetext{d}{{Lines} clearly detected when integrating over both
  components $7_0-6_0$ and $7_{-1}-6_{-1}$ components.}
\tablenotetext{e}{Compact, red-shifted emission observed toward one of
  the binary components IRAS4A-NW (see discussion in text).}
\end{deluxetable}

As with the continuum observations, a significant fraction of the line
emission is resolved out - and only the brightest, most compact
regions are detected. For most species and sources, this occurs close
to the continuum positions - as one would expect from a centrally
condensed envelope. It is also seen that the sources which are
observed with the shortest possible baselines in the Compact
configuration show stronger emission in some of the prominent lines:
for example, strong C$^{17}$O 3--2 emission is detected in L1157
whereas it does not show particularly strong C$^{17}$O 3--2 lines
compared to the other sources in the survey of \cite{jorgensen02}.
Again, this likely reflects that most of the C$^{17}$O 3--2 emission
seen in the single-dish observations comes from the extended envelope
and thus is very sensitive to the differences in the lengths of the
shortest baselines.

The amount of resolving out for a somewhat extended molecular species
is estimated by comparison to the single-dish observations of
C$^{18}$O 2--1 by \cite{jorgensen02} (for B335, CSO observations
presented by \cite{evans05b335} were used). The integrated line
intensities from the single-dish observations are converted from
antenna temperature scales to Jy~beam$^{-1}$ using the standard
expressions and compared to the central spectra from either the
interferometer datasets restored with a beam size comparable to the
single-dish beam size (in cases where the signal-to-noise is
sufficient) or the flux integrated over a smaller region where the
emission was detected. For C$^{18}$O 2--1 the amount of flux recovered
by the interferometer estimated in this way varies from 1--3\% for
L1448, IRAS2A, IRAS4B, L1527 and L483, about 6\% for IRAS4A to 16\%
and 22\% for L1157 and B335, respectively. The slightly higher amount
of recovered emission in IRAS4A (compared to the other sources
observed in the Compact-North configuration) and vice versa the
slightly lower amount of recovered emission in L483 (compared to the
sources observed in the compact configuration) are nice examples of
the importance of source structure. IRAS4A has significant structure
within the single-dish beam, in particular by the two binary
components well separated by the interferometer and each contributing
to the total flux, whereas L483 typically shows a more smooth
distribution of its emission over larger scales. Still, since only
15--20\% of the emission is recovered even in the best cases of B335
and L1157, it is clear that care must be taken when interpreting the
interferometric data. It is, for example, not possible to deduce
quantitative information from the integrated line fluxes without
compensating for the lack of short spacing data.

\begin{figure}
\resizebox{\hsize}{!}{\includegraphics{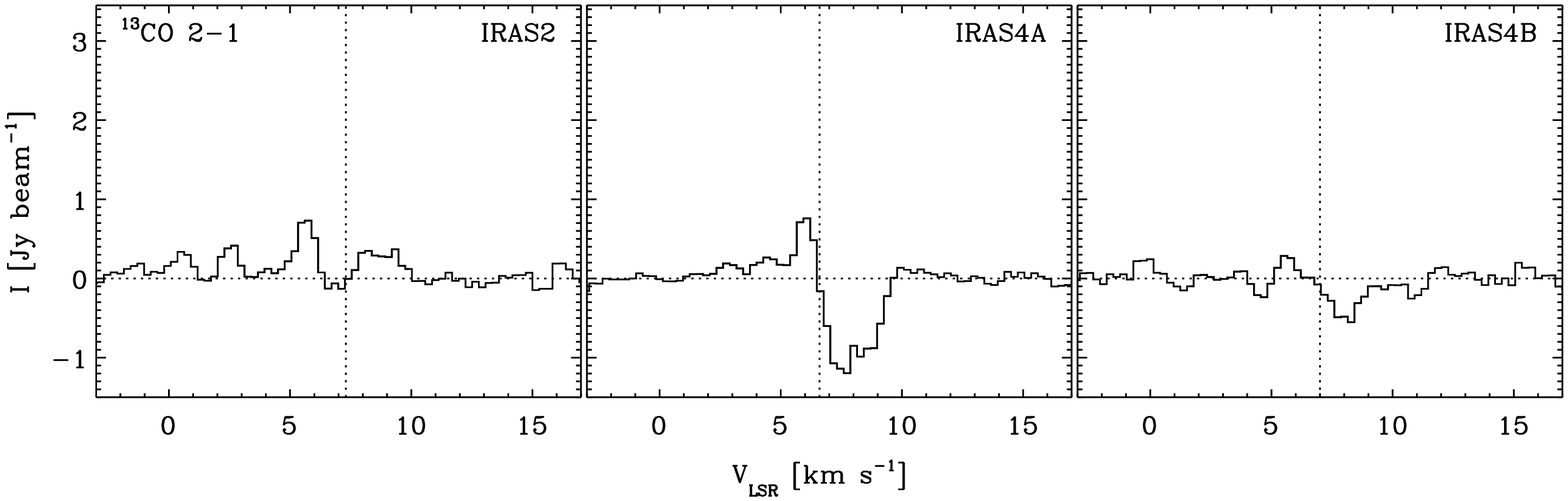}}
\caption{$^{13}$CO 2--1 spectra toward the center positions of IRAS2A,
  IRAS4A and IRAS4B.}\label{pcyg_spectra}
\end{figure}

Two sources, IRAS4A and IRAS4B, show characteristic ``inverse P
Cygni'' line profiles where the red-shifted part of the lines is seen
in absorption against the continuum while the blue-shifted part is
seen in emission. Such profiles can be taken as the least ambiguous
evidence for infall in protostellar cores where the foreground
material falling toward the central source (red-shifted) is absorbing
the continuum emission. Inverse P Cygni profiles have been seen
previously in lines of H$_2$CO and CS toward IRAS4A and IRAS4B in IRAM
Plateau de Bure observations by \cite{difrancesco01}. With the SMA
operating at submillimeter wavelengths (and therefore measuring
stronger continuum emission and high excitation lines), it should be
ideally suited for detecting such line profiles. Of all the lines in
the sample, only the $^{13}$CO 2--1 lines toward IRAS4A and IRAS4B
show clear inverse P Cygni profiles
(Figure~\ref{pcyg_spectra}). Together with IRAS~16293-2422
\citep{chandler05,takakuwa06,remijan06}, these sources provide the
only known examples of inverse P Cygni profiles toward low-mass
protostars. It could suggest that the observed inverse P Cygni
profiles are related to much larger scale infalling motions such as
those seen in single-dish observations of IRAS4 by \cite{walsh06}
rather than collapse onto the central protostar. The 230~GHz continuum
emission toward IRAS4A is marginally resolved between the two
components of the binary and no significant difference is seen between
the widths of the absorption across the continuum possibly arguing in
favor of the presence of more global collapsing motions.
\clearpage

\clearpage
\centerline{\resizebox{!}{0.9\vsize}{\includegraphics{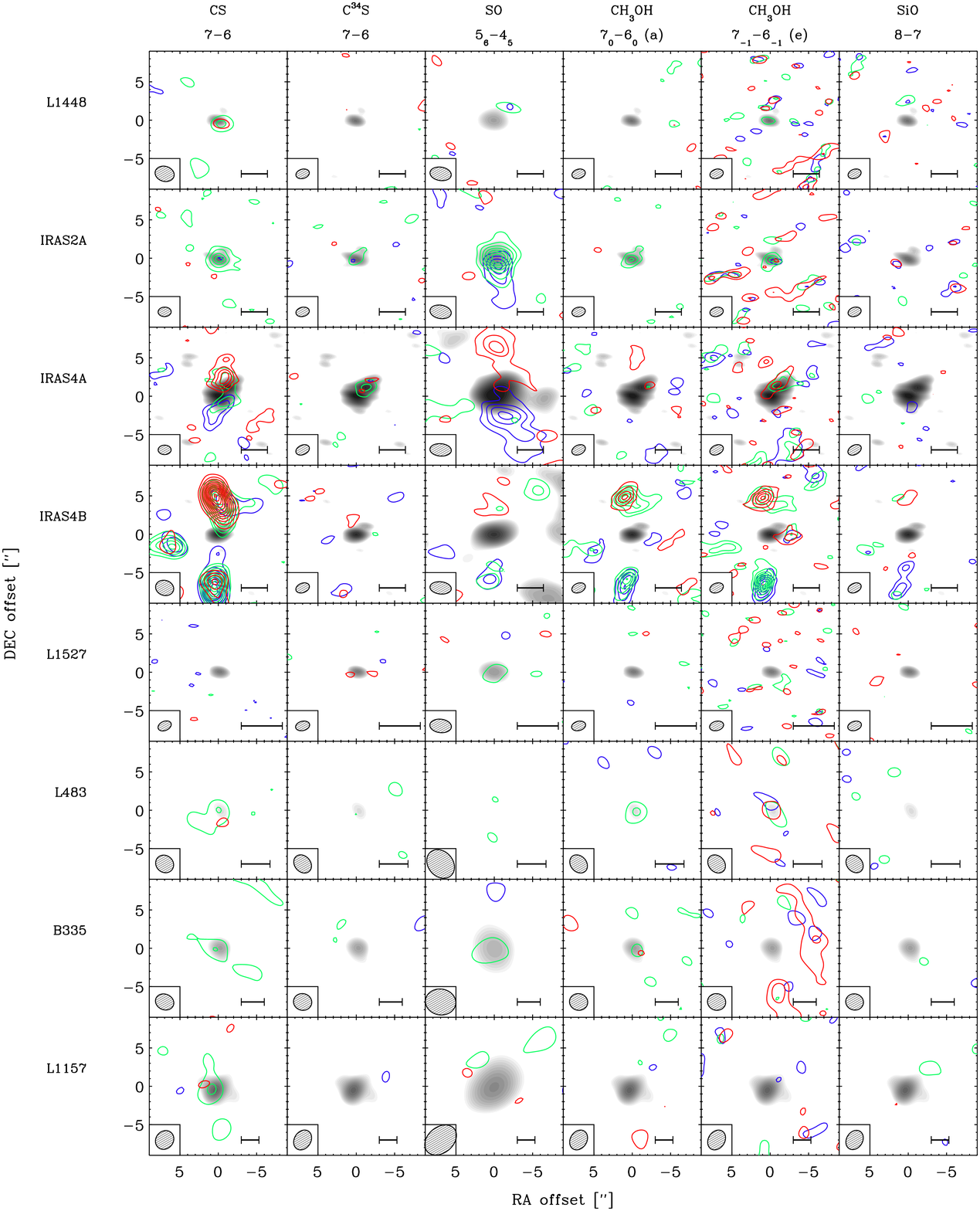}}}
\clearpage
\clearpage

\subsection{Dynamical impact of outflows}\label{cooutflow}
\begin{figure}
\resizebox{\hsize}{!}{\includegraphics{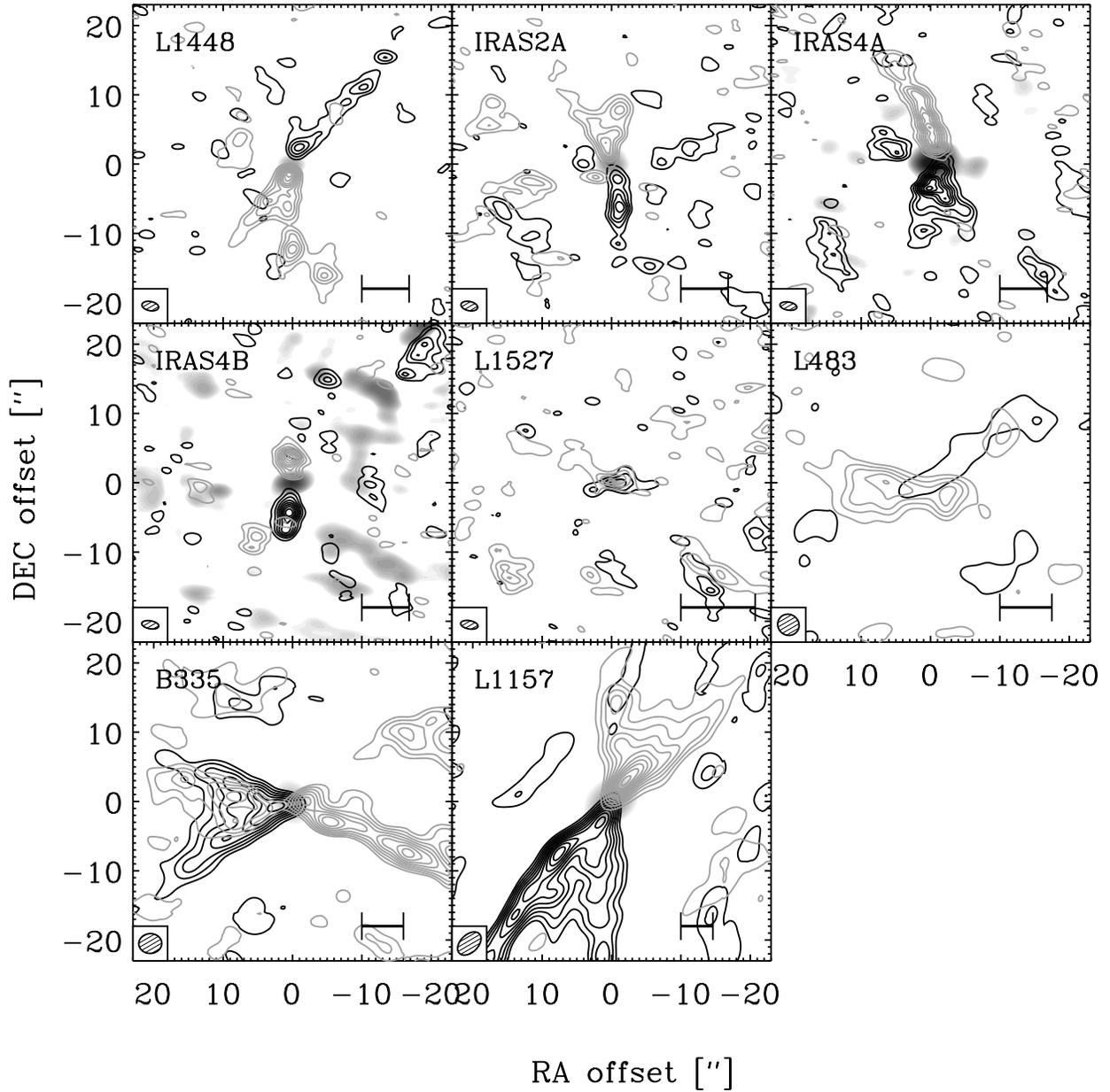}}
\caption{Map of the CO 2--1 emission over the entire interferometer
  field of view for each source in the sample. The contours are shown
  in steps of 3$\sigma$ with the black contours indicating emission
  integrated from $-$6~\kms\ to $-$1~\kms\ relative to the systemic
  velocity and grey contours emission integrated from +1~\kms\ to +6~\kms\
  relative to the systemic velocity. The scale bar in the lower right
  corner of each plot indicates a length of 1500~AU at the distance of the
  respective source.}\label{co_maps}
\end{figure}

Of all the lines in the survey, the CO 2--1 lines provide the most
striking features, showing extended outflow emission over most of the
SMA primary beam (Figure~\ref{co_maps}). The CO emission has
significant differences, from the highly collimated IRAS4B outflow to
the somewhat more diffuse L483, L1527 and B335 outflows. The close
binarity of IRAS2A and IRAS4A is also seen in the CO maps causing more
confusing pictures of those outflows. IRAS2A for example shows hints
of both a prominent north-south cone-shaped outflow and a more
collimated east-west jet \citep[see also discussion
in][]{i2art,n1333i2art}. For L1448, IRAS2A and L1157, the CO emission
is seen to delineate the edges of the outflow cavities.

It should be emphasized that some of the differences between outflows
in this sample could be due to inclination effects. A highly
collimated outflow directed in the plane of the sky will have small
relative CO velocities over the extent of the outflow and its CO
emission will be more susceptible to resolving out as it merges with
the surrounding cloud material. L1527 shows a ``butterfly'' shape in
lower-$J$ CO and HCO$^+$ lines \citep{tamura96,hogerheijde97},
suggesting that inclination is significant there. Likewise, an outflow
with a large opening angle viewed at an inclination close to the
opening of the outflow cones will show both red- and blue-shifted
emission overlapping in both lobes, which has been shown to be the
case for B335 \citep{cabrit88}.

Recently, \cite{arce06} imaged a sample of nine Class 0, I and II
young stellar objects in the $^{12}$CO 1-0 line observed with the
Owens Valley Radio Observatory Millimeter Array. They found that
outflow opening angles are a good discriminant between the different
evolutionary stages of the objects and suggested that it reflects that
the material in the protostellar envelope is cleared as the outflow
opens up an increasingly wide cavity. A similar suggestion has been
put forward by \cite{velusamy98}.

All of the outflows in our sample are found to have small opening
angles ($\lesssim 60^\circ$), further supporting the conclusion of
\citeauthor{arce06} that the outflows with the narrowest opening
angles belong to the Class 0 objects. The individual outflows in our
sample share many of the features with the three different Class 0
outflows in the sample of \citeauthor{arce06}, ranging from the
jet-like IRAS4B, IRAS4A2 and IRAS2A (east-west) outflows (similar to
the HH114-mms outflow from the sample of \citeauthor{arce06}), to the
more cone-like L1448, IRAS2A (north-south), B335 and L1157 outflows
(similar to the IRAS~03282+3035 outflow in the \citeauthor{arce06}
sample). In addition the diffuse L483 and L1527 outflows appear more
similar to the Class I outflows in the sample of
\citeauthor{arce06}. It is interesting to note that no significant
changes in the shape of the cone is seen for the observed outflows all
the way to the few hundred AU scales probed by the SMA observations,
in contrast with what is expected if the outflow angle increases over
the duration of the Class 0 stage. That is, the outflows do not show a
parabolic shape reflecting a larger opening angle close to the central
protostar compared to material further away - in contrast to what is
seen in the perhaps best example of this scenario in B5-IRS1
\citep{velusamy98}.

It is also interesting to note that the most collimated outflows
(IRAS4A and IRAS4B) are related to the objects with the highest ratios
of disk to envelope masses from the continuum studies while the
objects with the lowest disk masses compared to envelope masses (L1527
and L483) show the most diffuse outflows. This may seem a disparity;
in a simple picture of envelope material falling into a circumstellar
disk and accreting onto the central protostar, the mass of the disk is
expected to increase until the envelope has dissipated
\citep[e.g.,][]{nakamoto94,hueso05}. One would therefore expect that
the more evolved (still embedded) objects would have the highest
disk-to-envelope mass ratios in contrast to what is observed here.

The question is whether L483 and L1527 still will accrete significant
material from the currently observed dust envelope. As pointed out in
\S\ref{cont_morph}, both show very flat density profiles from the
SCUBA maps, which could reflect the importance of heating of the
protostellar envelopes through their outflow cavities. Indeed, both
sources show near-infrared scattering nebulosities
\citep[e.g.,][]{tamura91,fuller95,l483art}, supporting the view that
their envelopes are being heated at larger scales. The high angular
resolution N$_2$H$^+$ observations of L483 by \cite{l483art}
furthermore indicate a velocity gradient in the larger scale quiescent
envelope material in the direction of the outflow, likely reflecting
that the envelope indeed is being dispersed by the action of the
outflow. It is therefore plausible that L1527 and L483 in particular,
are in a phase where the accretion has more or less come to a stop -
and the protostar just is ``waiting'' to disperse its remaining
envelope through the action of the outflow. The relatively large disk
masses of the remaining objects could imply that they also are close
to the maximum disk masses - but comparisons to both disk and envelope
structures of more evolved Class I objects are needed to confirm or
refute this. Also, firmer constraints on both disk structures and
outflow parameters for these sources will be important to obtain a
more quantitative picture of the dynamical properties of the envelope
and disk material.
\clearpage

\subsection{Chemical impact of outflows}
Another example of the importance of the outflows is most clearly seen
in the chemical differentiation toward some of the sources. For
IRAS4B, for example, a number of lines show emission separated into
clumps about 6\arcsec\ (1500~AU) north/south of the central protostar,
prominently seen in CS, CH$_3$OH and H$_2$CO. These species are found
to be located on the tips of the CO outflows and likely probe shocks
where the outflow impacts the envelope. In IRAS4A, characteristic
shocks (although less prominent than in IRAS4B) are seen north and
south in the outflow associated with the northwestern binary
component, IRAS4A-NW (IRAS4A2). Fainter compact emission is also seen
toward the center of this source; interestingly, this emission appears
to be red-shifted with respect to the systemic velocity, also
suggesting an outflow impact on the smallest scales. The shock-tracing
SiO and SO emission also peak away from the central protostars in
IRAS4A and IRAS4B, although only a very tentative SiO emission is
observed toward IRAS4A. For three other sources in the sample, compact
CH$_3$OH emission is seen, most prominently in IRAS2A. In contrast to
the two other sources in NGC~1333, a number of high excitation
transitions are observed in the CH$_3$OH $7_k-6_k$ band toward this
source. This is indicative of the presence of hot gas on small scales
in IRAS2A, consistent with the conclusion of \cite{iras2sma} who imaged
even higher excitation CH$_3$OH transitions also with the SMA.

The three sources in NGC~1333 (IRAS2A, IRAS4A and IRAS4B) have all
been suggested to have ``hot corinos'' \citep[e.g.][]{maret04}, i.e.,
inner regions where the temperature increases above 100~K due to
passive heating by the protostar and where complex organic molecules
are produced into the gas-phase as ice mantles evaporate. This was in
particular inferred from multi-transition observations of H$_2$CO and
CH$_3$OH which in some cases could not be fit by constant
abundances. \cite{maret04,maret05} discussed these single-dish
observations in context of the hot corino scenario and found that
these sources required an abundance enhancement by up to four orders
of magnitude in the innermost regions. \cite{hotcoresample} on the
other hand demonstrated that those abundance enhancements in most
cases were a result of specific assumptions, e.g., about the ratio
between the abundances ortho- and para-H$_2$CO. Furthermore, shocks
caused by the outflows can enhance the abundances of these species on
larger scales which further complicates the interpretation of hot
corinos: \cite{hotcoresample}, for example, found that the H$_2$CO and
CH$_3$OH abundance variations in IRAS4B were best explained by a
scenario in which the molecules were released in the outflow driven
shocks, further supported by the line profiles of these species and
high excitation CS 10--9 lines.

To properly address these issues high angular resolution observations
are clearly important, e.g., to directly image where molecular species
have their origin. \cite{mundy92} for example demonstrated that the
emission of SO toward the inner few arcseconds of the Class 0 source
IRAS~16293-2422 was enhanced by the action of the outflow also traced
by centimeter continuum emission and H$_2$O maser activity. More
recently, \cite{bottinelli04iras16293} resolved the emission of
complex organic species, CH$_3$CN and HCOOCH$_3$, toward this source
and derived a size of $\sim 100$~AU, comparable to the expected size
of the hot corino. \cite{chandler05} on the other hand found that the
emission of high excitation transitions and complex organic molecules
toward this source was shifted by a few tenths of an arcsecond ($\sim
50$~AU) coinciding with the shock seen in longer wavelength continuum
observations. The SMA observations presented in this paper indicates
that the peaks in the CH$_3$OH and H$_2$CO emission toward IRAS4A-NW
and IRAS4B are related to the shocks and that no trace of compact
emission from these species is seen toward the central continuum
sources, in contrast to what should be expected in the case of hot
corinos. We note that also for these transitions the SMA resolve out a
significant fraction (50--75\%) of the observed single-dish flux
indicating that H$_2$CO and CH$_3$OH are present in the large scale
envelopes.

IRAS2A still stands out with compact hot gas as witnessed, e.g., by
the detection of many of the higher excitation hyperfine components of
the $J=7-6$ line (Fig.~\ref{ch3oh_spectra} and higher $J$ CH$_3$OH
lines \citep{iras2sma} that appear unresolved. Of the well-studied
low-mass protostars, IRAS2 therefore appears to be the best hot corino
candidate. Given the importance of outflows in the other low-mass
sources where complex organic molecules have been observed, however,
it is a valid question to ask whether the protostellar outflow is
important on these small scales in IRAS2A, where it is still
unresolved by the SMA observations (similar to the case of
IRAS~16293-2422). The SO emission do show evidence of a velocity
gradient which could suggest that the shocks are impacting the
material on small scales toward IRAS2A. Also, it is possible that
accretion shocks in the disk could temporarily heat the gas in the
surface layer to high temperatures and likewise inject complex
molecules from the ice mantles into the gas-phase. In IRAS2A in
particular, this seems to be a feasible scenario, given the extent of
the disk from submillimeter continuum data \citep[see discussion
in][]{iras2sma}. Again, further high-resolution line observations
could potentially disentangle the velocity field of the disk and
further address its physical properties (temperature, density) and
chemical structure.

This discussion directly illustrates the importance of high-angular
resolution imaging for the interpretation of chemical variations in
protostellar environments. With the high-resolution SMA observations,
it becomes possible to image directly these physical and chemical
variations in the immediate vicinity of the low-mass protostars.

\begin{figure}
\resizebox{\hsize}{!}{\includegraphics{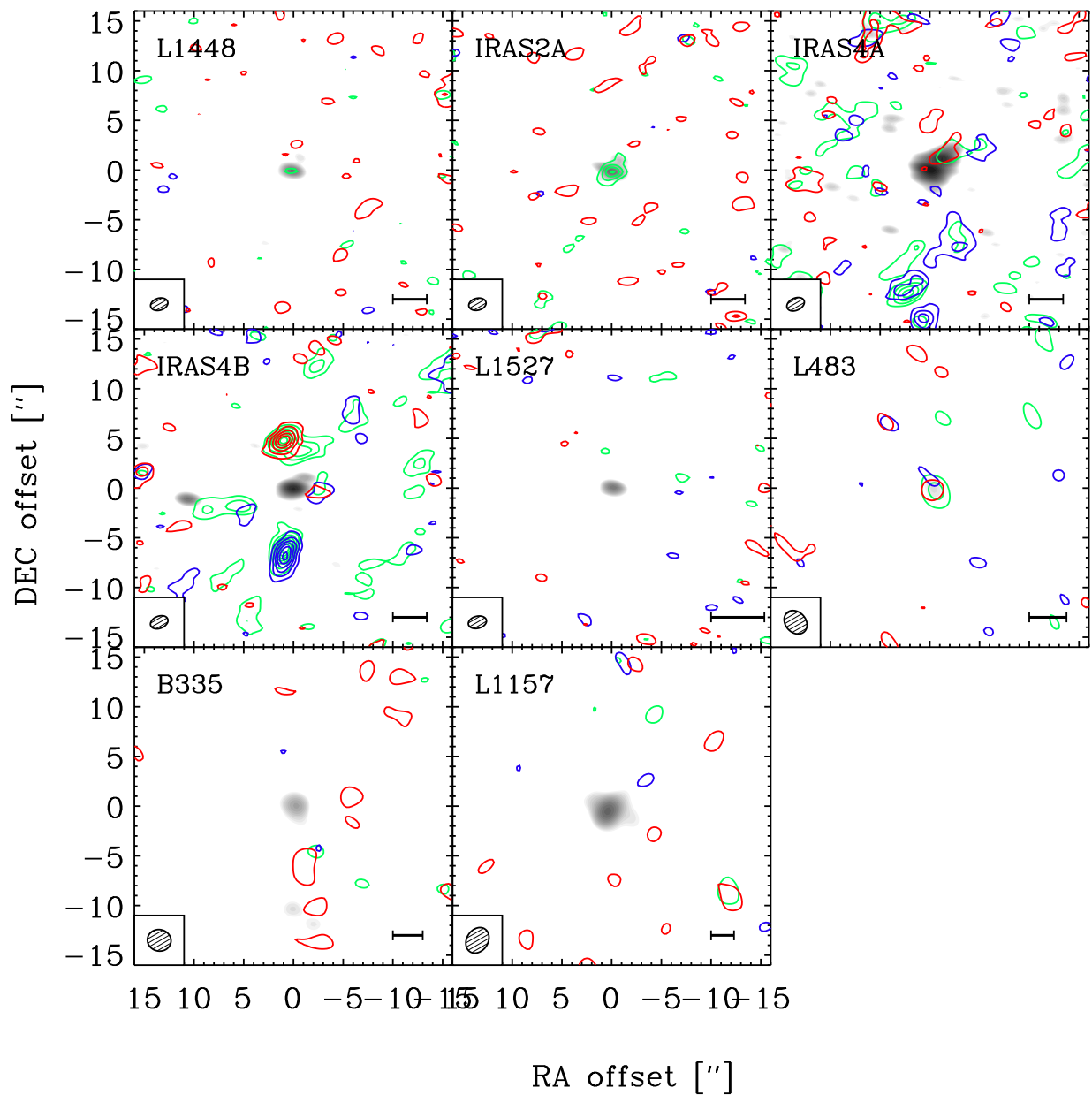}}
\caption{CH$_3$OH $7_0-6_0$ and $7_{-1}-6_{-1}$ emission (added-up)
  toward the entire sample of sources. The emission has been
  integrated over the same velocity intervals as in
  Fig.~\ref{firstmolmap}. The scalebar in the lower right corner of
  each plot indicates a scale of 750~AU at the distance of each
  source.}\label{ch3oh_molmap1}
\end{figure}
\begin{figure}
\resizebox{\hsize}{!}{\includegraphics{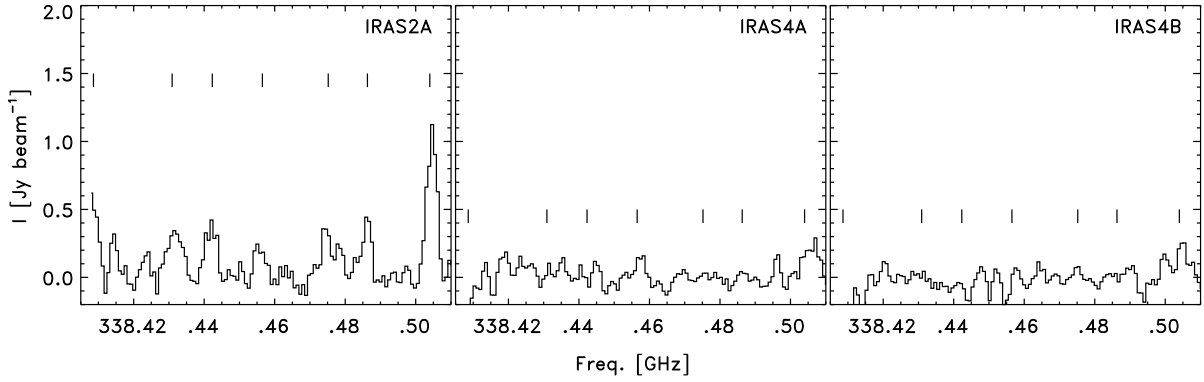}}
\caption{Higher excitation CH$_3$OH $7_k-6_k$ transitions toward the
  center positions of IRAS2A, IRAS4A and IRAS4B. The location of the
  transitions have been marked with vertical lines, from left through
  right: 0A (also shown in Figure~\ref{ch3oh_molmap1}), -6E, 6A, -5E,
  5E, 5A and -4E.}\label{ch3oh_spectra}
\end{figure}
\clearpage
\section{Summary}\label{summary}
A large line and continuum survey with the Submillimeter Array of 8
deeply embedded, low-mass protostellar cores has been presented. Each
source has been observed in three different spectral settings that
included lines of the most common molecular species, CO, HCO$^+$, CS,
SO, H$_2$CO, CH$_3$OH and SiO. Emission from (sub)millimeter lines
from 11 molecular species (including isotopologues) have been imaged
at high angular resolution (1$''$--3$''$; typically corresponding to
200--600~AU) together with continuum emission at 230 (1.3~mm) and
345~GHz (0.8~mm). A number of conclusions can be inferred from simple
considerations of this dataset:

\begin{itemize}
\item Compact continuum emission is observed for all sources.
  Comparing the 230~GHz and 345~GHz data, the emission is found to be
  well described as $F\propto \nu^\alpha$ with an average value of
  $\alpha$ of 2.4. $\alpha > 3.0$ was only found for one source,
  L483. For this source, the compact continuum emission is weak at
  230~GHz and the emission comes from the envelope rather than the
  circumstellar disk. For the remaining sources, the continuum
  emission likely originates in the disks being marginally optically
  thick and having dust opacities, $\kappa_\nu \propto \nu^\beta$ with
  $\beta \approx 1$. Lower limits to the total disk masses (gas+dust)
  are found to be typically 0.1~$M_\odot$ for a dust temperature of
  30~K (1--10\% of their envelope masses).
\item Inverse P Cygni profiles indicative of infalling motions are
  seen in the $^{13}$CO 2--1 lines toward IRAS4A and IRAS4B. These are
  not observed in other lines or toward other sources (besides
  IRAS~16293-2422), however, and it is possible that these profiles
  have their origin in the ambient cloud/outer envelope rather than
  infall in the inner envelope themselves.
\item Prominent outflows are present in all sources extending over
  most of the interferometer field of view. Most of the outflows show
  small opening angles $\lesssim 60^\circ$ comparable to other
  previously studied Class 0 objects - and in agreement with the
  suggestion that the outflow opening angle increases with time
  \citep{velusamy98,arce06}. Still, significant differences are seen
  in their morphologies with some outflows showing more jet-like
  structure and others seemingly tracing material in the outflow
  cavity walls. The most diffuse outflows are found for the objects
  with the lowest ratios of disk-to-envelope masses, in particular
  L1527 and L483, which are thought to be transitionary between the
  Class 0 and I stages. It is suggested that these sources have
  reached the end of their main accretion and that the remaining
  envelope material will be dispersed by the outflows.
\item Shocks are present on all scales in the protostellar
  environments, most clearly traced by extended emission of CH$_3$OH
  in IRAS4A and IRAS4B. Compact red-shifted CH$_3$OH emission is seen
  toward one of the components of the IRAS4A binary, IRAS4A-NW
  (IRAS4A2). These observations suggest that the emission of H$_2$CO
  and CH$_3$OH, often taken as signposts of low-mass ``hot corinos'',
  are in fact related to the shocks caused by the protostellar
  outflows. In this context, IRAS2A stands out as the only source with
  clear evidence for hot, compact gas (as also seen in previous SMA
  observations by \citealt{iras2sma}). Further high resolution
  observations will be necessary to ascertain whether this emission is
  related to a passively heated hot corino, a shock in the
  circumstellar disk or perhaps just the action of the outflow on the
  smallest scales.
\end{itemize}

It is clear that these high-resolution observations pushes our
understanding of embedded protostars as simple and quiescent
structures toward a more dynamic picture. Outflows play an important
role in all aspects of the structure of these low-mass protostellar
systems, from the larger scale swept up material clearly probed by the
CO outflows down to the smallest (arcsecond) scales where shocks are
present. It is important to realize that the larger scale envelope
carrying the bulk of the material probed by single-dish observations
is almost completely resolved out by the SMA observations. Using the
SMA data in conjunction with single-dish data and/or detailed models
for the source structures, it will become possible to make
quantitative statements, e.g., about the physical structure of the
envelopes or the chemical abundances from scales of thousands of AU
down to just the few hundred AU probed by the interferometers. More
detailed discussions of aspects of star formation based on the data
presented in this paper, in addition to such single-dish data and
detailed radiative transfer models, will be presented in a series of
upcoming specialized papers.

\acknowledgements It is a pleasure to thank everybody involved with
the Submillimeter Array for the continued development of this
instrument. We are also grateful to the observers who have contributed
by observing one or more tracks in this program. The research of
J.K.J. was supported by NASA Origins Grant
NAG5-13050. F.L.S. acknowledges support from the Swedish Research
Council. Astrochemistry in Leiden is supported by a NWO Spinoza grant
and a NOVA grant.

\appendix
\section{Description of sources}\label{sampledescription}
\subsection{L1448-C(N)}
L1448-C(N) is one of $\sim$10 young stellar objects in the L1448
complex in the Perseus molecular cloud. Recent Spitzer observations
\citep{perspitz} have resolved it into two components with a
separation of about 8\arcsec\ (2000~AU). The northern of these two is
known from millimeter aperture synthesis observations
\citep[e.g.,][]{barsony98,hotcorepaper} which is taken as the pointing
center for the observations in this paper.

\subsection{NGC~1333-IRAS2A}
NGC~1333-IRAS2A is part of a protostellar system also encompassing the
protostar IRAS2B and likely pre-stellar core IRAS2C seen in SCUBA maps
\citep{sandell01}. Both IRAS2A and IRAS2B have been imaged at high
angular resolution at millimeter wavelengths
\citep{blake96,looney00,n1333i2art} whereas IRAS2C does not show any
compact continuum emission. IRAS2B is located outside the primary beam
field of view at a separation of 30$''$ (7500~AU), but is detected in
the 1.3~mm maps presented here. IRAS2A itself has been speculated to
be a protostellar binary, resulting in its characteristic quadrupolar
outflow (see also \S\ref{cooutflow}). At 0.8~mm, it appears to be
resolved into two separate components with a fainter companion
northeast of the main peak, but as for L1448 above, it is not clear
whether this is simply a result of resolving out differing amounts of
envelope material. \cite{iras2sma} examined SMA continuum observations
from IRAS2A and showed that the emission could be explained by a
combination of an extended envelope and a 300~AU diameter disk.

\subsection{NGC~1333-IRAS4A}
NGC~1333-IRAS4A (together with its companion IRAS4B; both
well-separated in SCUBA maps) is one of the most well-studied
protostellar systems in NGC~1333. \cite{difrancesco01} detected
inverse P~Cygni profiles indicative of infalling motions in lines of
H$_2$CO toward these sources. IRAS4A is resolved into two binary
components, NGC~1333-IRAS4A-NW and -IRAS4A-SE with a separation of
1.8$''$ (400~AU) \citep{lay95,looney00,difrancesco01}.

\subsection{NGC~1333-IRAS4B}
NGC~1333-IRAS4B is located about 30\arcsec\ southeast of
NGC~1333-IRAS4A and also shows signs of inverse P~Cygni profiles
\citep{difrancesco01}. A fainter companion NGC~1333-IRAS4B$'$ (also
sometimes referred to as NGC~1333-IRAS4C or NGC~1333-IRAS4B2) is
located another 10\arcsec\ to the east
\citep{looney00,difrancesco01}. The outflow in IRAS4B appears to play
an important role of 5--10\arcsec\ scales affecting the emission of a
number of molecular species. \cite{hotcoresample} for example found
that the abundance variations of H$_2$CO and CH$_3$OH deduced from
single-dish observations best were explained by the impact of this
outflow - also accounting for the line profiles of these species and
high excitation CS 10--9 emission detected toward this source.

\subsection{L1527-IRS}
L1527-IRS (IRAS~04368+2557) is one of the possibly more evolved
protostars in the sample located in a dense core in Taurus. L1527-IRS
was previously studied at high angular resolution with the Nobeyama
Millimeter Array \citep{ohashi97} and the Owens Valley Radio
Observatory Millimeter Array
\citep[OVRO;][]{hogerheijde97,hogerheijde98} and found to show a
characteristic X-shaped structure in lines of $^{13}$CO 1--0 and
HCO$^{+}$ that likely delineates the walls of an outflow cavity. It
was suggested to be a binary based on 800~$\mu$m continuum
observations \citep{fuller96} with the secondary at about 21\arcsec\
from the main protostar. No evidence is found of this secondary in the
previous high resolution 3~mm maps or in the maps presented here,
however (it would be located outside the primary beam field of view at
345~GHz, but inside at 230~GHz). The SCUBA maps show a very extended
core with little central concentration (a flat envelope density
profile $n \propto r^{-p}$ with $p=0.6-1.0$ is found for this source
\citep{jorgensen02,shirley02} whereas most of the other protostars
have steeper profiles with $p=1.5-2.0$).

\subsection{L483-mm}
L483-mm is a strong infrared source (IRAS~18148-0440) associated with
an isolated core. \cite{tafalla00} suggested it was transitionary
between the Class 0 and I stages. It has been imaged at high
resolution in continuum and lines at 3~mm with the Berkeley Maryland
Illinois Association (BIMA) Millimeter Array \citep{park00} and OVRO
MMA \citep{l483art}. In the analysis of the continuum emission
\cite{l483art} found that all the detected continuum emission could be
attributed to the extended envelope, contrasting L483-FIR from L1448-C
and NGC~1333-IRAS2A described above where a compact continuum source
was present. \cite{takakuwa06aste} recently mapped L483-mm in the HCN
4--3 transition at submillimeter wavelengths using the ASTE
single-dish telescope. They found that the HCN emission was slightly
extended, which they argued is reflecting the heating of the envelope
material to 4000 AU scales through the outflow cavity.

\subsection{B335}
B335 has often been considered one of the best protostellar collapse
candidates. It has been imaged in continuum at subarcsecond resolution
\citep{harvey03,harvey03b} and in lines
\citep{wilner00}. \cite{harvey03b} found that the continuum emission
could be well described by a $n \propto r^{-1.5}$ envelope down to
scales of $\sim 100$~AU within which a disk with a size
$\lesssim$~100~AU is seen. \cite{wilner00} showed that the line
emission thought to trace the collapse were in fact dominated by a
number of clumps extending along the outflow cavity into the low
density medium.

\subsection{L1157-mm}
L1157-mm is perhaps best known for its prominent outflow
\citep{bachiller97,bachiller01}, but also its small-scale
envelope/disk structure has gotten a fair amount of attention.
\cite{gueth97} observed L1157 in lines of CO (and isotopologues) and
continuum using the IRAM Plateau de Bure interferometer (PdBI). They
found that the continuum emission probed a combination of a large disk
and an extended envelope with a clear outflow cavity whereas the CO
isotopologues probed a progression from the envelope, throughout its
cavities to the outflow itself. \cite{goldsmith99} and
\cite{velusamy02} observed transitions of CH$_3$OH at 3~mm and 1.3~mm
using OVRO, suggesting that these CH$_3$OH transitions had their
origin in a warm layer of the disk heated by an accretion shock.

\end{document}